\documentclass[reprint, superscriptaddress, amsmath, amssymb, aps, prb]{revtex4-2}

\usepackage{braket}
\usepackage{times}
\usepackage{graphicx}
\usepackage{amsfonts}
\usepackage{amsmath}
\usepackage{bm}
\usepackage{xcolor}
\usepackage{hyperref}

\usepackage{xcolor}

\newcommand{\old}[1]{}
\newcommand{\new}[1]{ #1}
\newcommand{\Renyi}{R\'enyi}
\newcommand{\err}[1]{\sigma^{\bm #1}}
\newcommand{\ssre}{second stabilizer \Renyi\, entropy}

\newcommand{\rcme}{\Renyi\, conditional measurement entropy}
\newcommand{\expct}[1]{\left\langle #1 \right\rangle}
\DeclareMathOperator{\tr}{tr}
\DeclareMathOperator*{\argmin}{arg\,min}

\definecolor{dkgray}{gray}{0.4}
\usepackage{ulem}

\usepackage{verbatim}
\definecolor{dkblue}{rgb}{0,0,0.5} 

\begin{document} 


\title{Phase transition in magic with random quantum circuits}


\author { Pradeep Niroula}
\email{pniroula@umd.edu}
\affiliation{Joint Center for Quantum Information and Computer Science, University of Maryland and NIST, College Park, MD  20742}
\affiliation{Joint Quantum Institute, University of Maryland and NIST, College Park, MD  20742}

\author { Christopher David White }
\affiliation{Joint Center for Quantum Information and Computer Science, University of Maryland and NIST, College Park, MD  20742}
\author{ Qingfeng Wang }
\affiliation{Chemical Physics Program and Institute for Physical Science and Technology, University of Maryland, College Park, MD  20742}
\affiliation{Joint Quantum Institute, University of Maryland and NIST, College Park, MD  20742}

\author{ Sonika Johri }
\affiliation{IonQ, Inc., College Park, MD  20740}

\author{ Daiwei Zhu }
\affiliation{IonQ, Inc., College Park, MD  20740}

\author{ Christopher Monroe}
\affiliation{Joint Center for Quantum Information and Computer Science, University of Maryland and NIST, College Park, MD  20742}
\affiliation{Joint Quantum Institute, University of Maryland and NIST, College Park, MD  20742}
\affiliation{IonQ, Inc., College Park, MD  20740}
\affiliation{Duke Quantum Center, Department of Electrical \& Computer Engineering \\ and Department of Physics, Duke University, Durham, NC  27701}

\author{Crystal Noel}
\affiliation{Duke Quantum Center, Department of Electrical \& Computer Engineering \\ and Department of Physics, Duke University, Durham, NC  27701}

\author{Michael J. Gullans}
\email{mgullans@umd.edu}
\affiliation{Joint Center for Quantum Information and Computer Science, University of Maryland and NIST, College Park, MD  20742}



\begin{abstract}
Magic is a property of quantum states that enables universal fault-tolerant quantum computing using simple sets of gate operations. 
Understanding the mechanisms by which magic is created or destroyed is, therefore, a crucial step towards efficient and practical fault-tolerant computation.
We observe that a random stabilizer code subject to coherent errors exhibits a phase transition in magic, which we characterize through analytic, numeric and experimental probes. 
Below a critical error rate, stabilizer syndrome
measurements remove the accumulated magic in the circuit, effectively protecting against coherent errors;
above the critical error rate syndrome measurements concentrate magic.  A better understanding of such rich behavior in the resource theory of magic could shed more light on origins of quantum speedup and pave pathways for more efficient magic state generation. 
\end{abstract}

\maketitle


A central goal in  physics and computer science is to understand the origins of possible computational speedups of quantum information processors over their classical counterparts.  Entanglement is a central resource for fault-tolerant quantum computing, but it is not necessarily sufficient to realize computational speedups. The notion of entanglement must be extended to distinguish between the production of ``easy’’ and ``hard’’ quantum states by fault-tolerant operations.
Notably, even when the quantum state of the processor is highly entangled,
computations consisting of only Clifford gates
---a finite, non-universal subgroup of the unitary group ---%
applied to stabilizer states, or eigenstates of Pauli operators,
can be efficiently simulated on classical computers \cite{gottesman_heisenberg,aaronsonImprovedSimulationStabilizer2004}. 
Non-stabilizer input states or non-Clifford gates, by contrast, are believed to be exponentially difficult to simulate on classical computers \cite{bravyiSimulationQuantumCircuits2018,buEfficientClassicalSimulation2019b}. 
On quantum computers non-Clifford gates are easy, however, in the context of error-corrected quantum computer, these states and operations still require costly magic state distillation or other gate-intensive protocols \cite{bravyi_universal_2005,fowlerSurfaceCodesPractical2012,ogormanQuantumComputationRealistic2017,campbellRoadsFaulttolerantUniversal2017c}.

A resource theory of stabilizer computation has emerged \cite{veitch_resource_2014} to study this division between easy (Clifford) and hard (non-Clifford) gates. In this theory, \textit{magic} is the resource that enables universal quantum computation; the amount of magic in a state determines how useful it is as a non-stabilizer input state in fault-tolerant synthesis of non-Clifford operations.
Magic has been used to bound quantum complexities \cite{buComplexityQuantumCircuits2022}
and to constrain tensor network models of AdS-CFT \cite{whiteConformalFieldTheories2021}.
Magic-generating non-Clifford operations have also been shown to be necessary for simulating quantum chaos \cite{leone2021quantum}.
Understanding the mechanisms by which magic can be generated or suppressed in a quantum circuit is, therefore, necessary not only to accelerate progress towards universal quantum computing but also to understand the limits in which quantum computations become classically accessible. 

A related aspect of quantum entanglement is its behavior
in monitored quantum circuits, such a as measurement-induced entanglement phase transitions \cite{skinner_measurement-induced_2018, li_measurement-driven_2019}.
Monitored quantum circuits consist of local gates (or time evolution),
interspersed with some rate or density of projective measurements.
The simplest example of a monitored quantum circuit is the error correcting code:
the state undergoes a series of entangling ``encoding" unitaries, followed by projective syndrome measurement
and final logical ``decoding" unitaries \cite{terhal2015}.
In general, monitored quantum circuits can display a \textit{measurement-induced phase transition in entanglement}.
These systems display evidence of a complicated phase diagram determined by the details of the circuit \cite{potter_entanglement_2022,fisher_random_2022,zabalo_operator_2021},
and have connections to percolation theory \cite{skinner_measurement-induced_2018,iaconis_measurement-induced_2020},
the theory of stabilizer codes \cite{gullans_dynamical_2020},
and statistical mechanics models \cite{bao2020theory,jian_measurement-induced_2020,li_statistical_2021,barratt_field_2022}. Such hybrid circuits have also been shown to exhibit related phase transitions beyond entanglement \cite{agrawal2022entanglement}.

\begin{figure*}
    \centering
    \includegraphics[width=\textwidth]{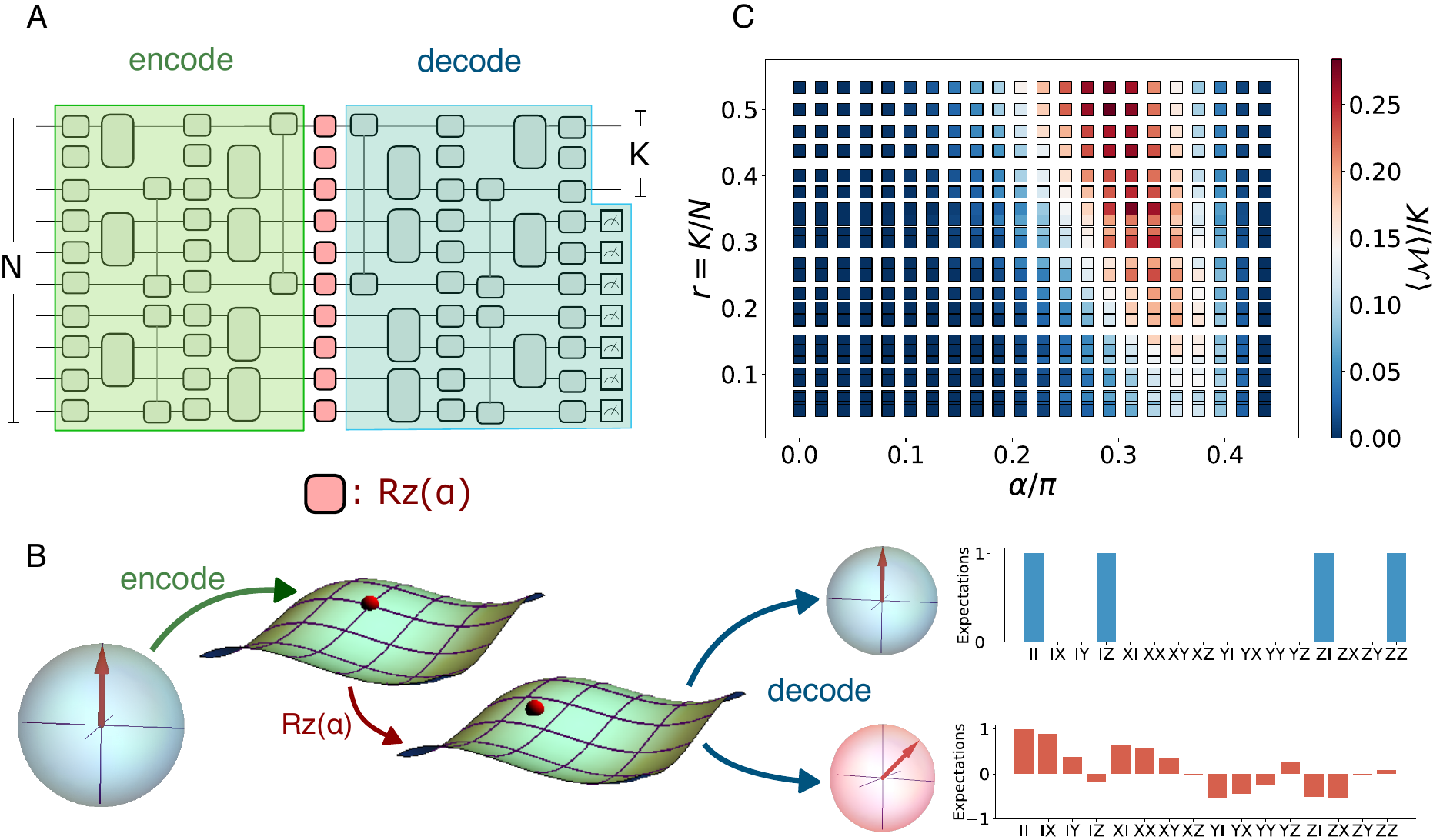}
    \caption{\textbf{Model and phase diagram}. \textbf{A}: The model. The qubits start in an all-zero state, corresponding to a logical $0$ state.
    We apply a random Clifford encoding circuit (green),
    controlled ``error'' unitaries (red),
    and the conjugate of the encoding circuit (blue).
    \textbf{B}: A schematic illustration of how magic is created or destroyed in our model. The encoding step acting on an input stabilizer state (represented by a blue Bloch sphere) produces a highly entangled stabilizer state in the many-qubit Hilbert space. Coherent rotations move the state off the grid of stabilizer states. The decoding step either snaps the state back to the grid of stabilizer states
    or pushes the state away from that grid. The final state is either a multi-qubit stabilizer state, represented by a Bloch-sphere shaded blue, or a magical state, represented by a Bloch sphere shaded red. The Pauli expectations of the resulting stabilizer (magical) state are shown as histograms shaded blue (red).
    \textbf{C}: Phase diagram for constant-rate codes. The color bar represents the magic density at a particular code-rate $r$, given by the ratio of logical qubits $K$ and total number of qubits $N$, and error rate, defined to be the angle of coherent rotation, $\alpha$.
    }
    \label{fig:overview}
\end{figure*}

In this paper, we show that measurement-induced phase transitions of entanglement can be extended to magic, and we study the transition experimentally. 
A quantum error correcting code subject to coherent errors displays a phase transition in the magic
as a function of the number of logical qubits (which in our model sets the measurement rate) or the error rate. 
In the magic phase transition, syndrome measurements, which can destroy magic, compete with errors, which can create magic, just as local dynamics and local measurement compete in the entanglement transition.
For large error rate or infrequent measurements,
the encoded state has extensive magic,
while for low error rate or frequent measurements,
the encoded state has nearly zero magic;
the two regimes are separated by a phase transition. A brief overview of our setup is sketched in Fig.~\ref{fig:overview}A and B and the resulting phase diagram is given in Fig.~\ref{fig:overview}C.
We also introduce a new measure of magic, the basis-minimized measurement entropy.
We measure this quantity and another known measure of magic, the stabilizer R\'enyi entropy \cite{leone_stabilizer_2022},
in  classical simulations, analytical calculations, and experiments on IonQ's Aria trapped-ion quantum computer. 
The magic phase transition is visible as a finite-size scaling collapse in these measures.

\section{Model} 
We study magic in random Clifford codes.
The initial state is a product state of $N$ qubits, $\ket{0}^{\otimes N}$. 
A randomly drawn Clifford circuit $C$ is applied to this state.
This Clifford circuit maps the initial state to the logical space of a random Clifford code; such codes are known to make high-performing error correcting codes \cite{brown2013short}.
After the encoding circuit $C$, a single-qubit rotation $R_z(\alpha) = \exp(-i\sigma_z \alpha/2)$ is applied to each qubit.
This ``noise layer'' models coherent noise and takes the quantum state away from the codespace of the Clifford code.
We call $\alpha$ the \textit{error rate}.
The noise layer is followed by $C^{\dagger}$, the conjugate of the encoding circuit.
Finally, $N-K$ qubits are measured in the computational basis, leaving a logical state with $K$ qubits.
These $N-K$ measurements are syndrome measurements for the Clifford code.

The encoding Clifford circuits are generated by interweaving $d$ layers of single-qubit unitaries and $d$ layers of two-qubit Clifford unitaries (Fig.~\ref{fig:overview}A). The odd layers are single-qubit gates sampled uniformly from the 24 elements of the single-qubit Clifford group. The even layers consist of fixed-angle ($\pi/2$) entangling M\o{}lmer-S\o{}rensen gates, defined as $\text{MS}(\pi/2) = e^{i\pi \sigma_x\sigma_x/4}$, applied to $N/2$ randomly chosen disjoint pairs. The decoding circuit is the inverse of the encoder.
We take $d = N$ in numerics and $d = N/2$ in experiment to reduce the effects of noise.  Circuits with $d=N/2$ have a behavior similar to those with depth $d=N$. In Supplementary Material (SM) Section \ref{app:half-depth-circuits}, we present numerics on $d=N/2$ and $d=2N$ circuits.

Fig.~\ref{fig:overview}B illustrates how magic is created or destroyed in our model. The state begins as a logical stabilizer state. The Bloch sphere, shaded blue, represents a multi-qubit stabilizer state. The encoding step maps the state to a stabilizer state in a many-qubit Hilbert space, and error moves the state off the grid of stabilizer states. The decoding (conjugate of the encoding operator together with syndrome measurements) step either snaps the state back to the grid of stabilizer states  or pushes the state away from that grid; in either case it projects the state back to the logical space. The final state is either a multi-qubit stabilizer state, represented by a Bloch sphere shaded blue, or a magical state, represented by a Bloch sphere shaded red. The stabilizer-ness of a state is visible in the Pauli expectations. For a stabilizer state, the distribution of expectations is concentrated among the stabilizing Paulis, as shown in the histogram shaded blue for a representative two-qubit stabilizer state. For a Haar state, the distribution has support over all Paulis, as shown in the histogram shaded red for a representative two-qubit Haar state. 

We study this transition for two different code rates (The code rate is the ratio of the number of logical qubits to underlying physical qubits.) The first case, which we refer to as ``vanishing rate", has only one logical qubit, so the code rate $r = 1/N$ tends to zero for large $N$. The second case uses constant-rate codes with the scaling $K = rN$ logical qubits for a fixed code rate $r$.

\section{Quantifying magic}
Any measure of magic for pure states is a function of quantum states that is zero for stabilizer states and nonincreasing under Clifford unitaries. 
Measures of magic can also be used to quantify the non-Clifford resources required to prepare a state, and
 how useful it can be in synthesizing non-Clifford gates via magic state distillation and injection.
We consider two measures of magic:
the second stabilizer R\'enyi entropy \cite{leone_stabilizer_2022, haug2023stabilizer} 
and the basis-minimized measurement entropy.

The second stabilizer R\'enyi entropy (SSRE) measures how spread out the state's density matrix is
when expanded in the basis of Pauli operators.
 A key property of stabilizer states is that they are the common eigenstate of a maximal set of mutually commuting Pauli operators \cite{gottesman1997stabilizer}. As a result,
the stabilizer state's density matrix is only supported on those operators,
so it is maximally concentrated and the SSRE is zero.
A Haar state on $N$ qubits, by contrast, has approximately equal weight on all Pauli operators,
so it is nearly maximally spread out and the SSRE, defined as $M_2(\rho) = -\log \frac{1}{2^N}\sum_{P \in \mathcal{P}} \text{Tr}(\rho P)^4$ for $N$ qubits, is proportional to $N$.
The histograms in Fig.~\ref{fig:overview}B  illustrate the distribution of Pauli expectations for these two cases.

We also consider a second measure of magic, which we call \textit{basis-minimized measurement entropy}, defined as the entropy of the Born probability distribution of measurement outcomes, minimized over the finite set of possible stabilizer measurement bases. For instance, consider a two-qubit stabilizer state $\ket{00}$ which we can measure in arbitrary length-two Pauli bases, including $X_1X_2$ and $Z_1Z_2$. 
Measuring $X_1X_2$ will result in a Born probability distribution of four equally possible measurement outcomes $\ket{\pm \pm}$, giving an entropy of $2$. 
On the other hand, measuring in $Z_1Z_2$ results in only one outcome $\ket{00}$, giving an entropy of 0.
Minimizing the entropy over all possible measured bases, the resulting basis-minimized measurement entropy is 0 in this case.
We wish to compute this basis-minimized measurement entropy for the resulting logical state in our model---%
that is, the state on the logical qubits after encoding, noise, application of the inverse of the encoding circuit, and syndrome measurement. In SM Section \ref{app:basis-minimized-entropy}, we show that the basis-minimized entropy is a good measure of non-stabilizerness for pure states. It is zero for a stabilizer state, is non-increasing under Clifford unitaries, and is subadditive for product states, i.e. $f(\sigma\otimes \rho) \leq f(\sigma) + f(\rho)$. 

The basis-minimized measurement entropy of the logical state depends on the syndrome outcome. Averaging the entropy of the logical state over all syndromes $\bm s$ gives us the basis-minimized classical conditional entropy
$\min_B S_{{\bm l_B} |\bm s} = \min_B \left(S_{\bm l_B, \bm s} - S_{\bm s}\right)$, where $S_{\bm s}$ is the entropy of the distribution of syndromes,
$B$ is a stabilizer basis for the logical Hilbert space, and $l_B$ is the outcome of measurement in stabilizer basis $B$. 
Furthermore, the conditional entropy without any basis minimization serves as a good upper-bound in the non-magical phase. 
Below the code's error-correction threshold, the logical state is close to the initial computational basis state, so we expect the optimal basis to be the computational basis.
So, for small $\alpha$ in our model, we expect the optimal basis to be the computational basis, and the conditional entropy is close to its optimal value (after basis minimization). 
Furthermore, the R\'enyi analogue of the conditional entropy, $S_{\bm l_B, \bm s}^{(2)} - S_{\bm s}^{(2)}$ where $S_X^{(2)} = -\log \sum_{x \in X}p_x^2$ is the R\'enyi entropy of distribution $X$, is analytically approachable. 
We compute the conditional entropy in classical simulation and experiment, and the R\'enyi analogue in experiment and analytical calculations.

The conditional entropy of the logical state quantifies the uncertainty in the logical space given a syndrome measurement, and it directly bounds the ability of a decoder to recover encoded classical information from measurements of the logical qubits (see SM Section \ref{supp:decoder-breakdown}).
A decoder is a syndrome-dependent operation that corrects logical errors corresponding to the syndrome measured. While the basis-minimized conditional entropy measures the minimal uncertainty over all possible Clifford decoding operations, the conditional entropy without basis-minimization limits the decoder to measurements in the computational basis.

In our experiment, we measure these measures of magic as a function of the error rate $\alpha$, tuning it from $0$ to $\pi/2$.
At zero error $(\alpha = 0)$ and maximal error ($\alpha = \pi / 2$), both measures are identically zero, because in each case the state is a stabilizer state.
When $\alpha=0$, the noise layer acts as the identity operator, the encoding circuit $C$ is cancelled by the following $C^\dagger$, and the final state is the same as the input stabilizer state. 
When $\alpha = \pi/2$, the error operator $e^{-i\sigma_z\alpha /2 }$ is itself a Clifford gate,
so the magic is likewise zero.

\section{Magic in the vanishing rate code}

\begin{figure*}[!ht]
    \centering
    \includegraphics[width=\textwidth]{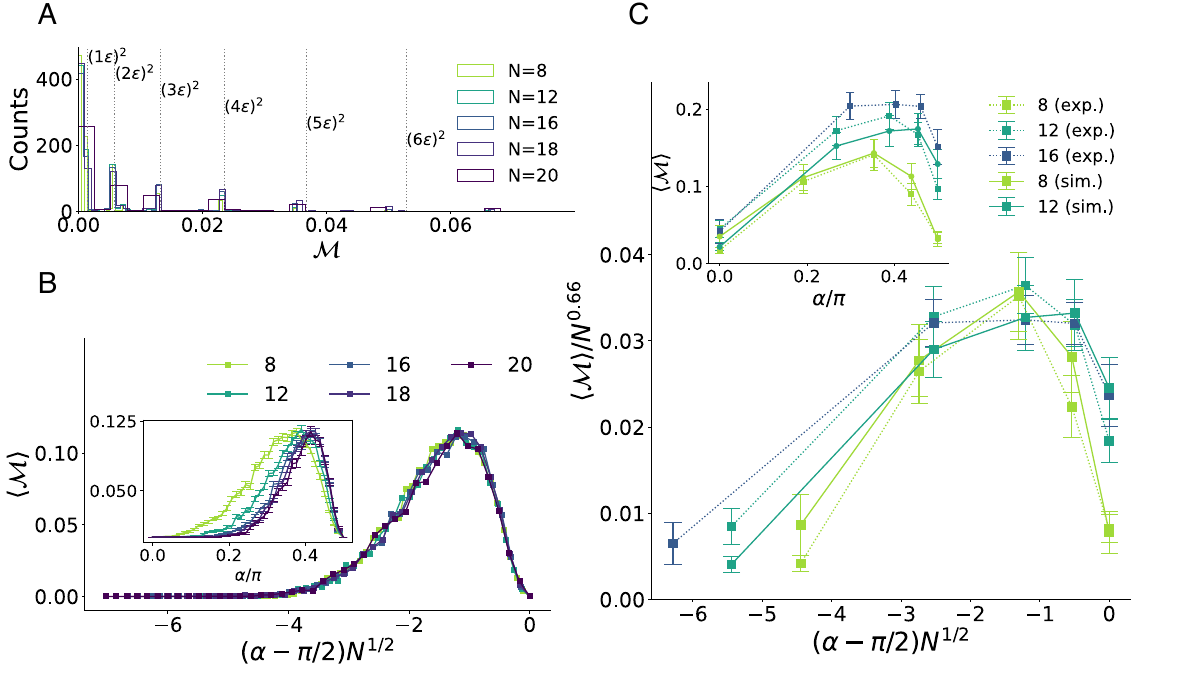}
    \caption{\textbf{Results for vanishing rate codes.}
    \textbf{A}: Distribution of second stabilizer R\'enyi entropy across codes and syndromes in classical simulation.
    The distribution is tightly peaked around square-integer multiples of the distance $\epsilon = \pi/2 - \alpha$ from the Clifford point, because it is controlled by the weights of the errors.
    \textbf{B \& C}: Syndrome- and circuit-averaged \ssre\, in classical numerics ({B}) and experiment on IonQ Aria trapped-ion quantum computer ({C}). Both display the predicted square-root scaling. The error estimates are derived using bootstrapping (details in the SM Section \ref{app:boostrapping-estimate}). The scaling with respect to system size of the vertical axis of C (main) is chosen to match the scaling of the peak in unscaled experimental data (inset).  For {B}, the errorbars are omitted in the collapse (main plot). In C, we also present numerics from noisy simulations (solid lines), obtained using a noise model that uses overrotation and depolarizing noise (See SM Section K for more details).
    }
    \label{fig:vanishing-rate}
\end{figure*}

First, we discuss the vanishing rate case with a single logical qubit. Between the two special Clifford points $\alpha = 0,\pi/2$ the  logical qubit has finite magic according to \ssre, with a peak at a distance $\propto 1/\sqrt{N}$ away from the Clifford point $\alpha=\pi/2$ point (see Fig.~\ref{fig:vanishing-rate}B). At large $N$ the Clifford point, therefore, becomes a singularity. 
We can understand the square root scaling by perturbing around the Clifford point $\alpha = \pi/2$.
At the Clifford point, the logical state is not magical
because it is an equal superposition over states corresponding to that syndrome.
Away from the Clifford point,
the logical state becomes magical to the extent the amplitudes in the superposition are no longer equal.
If exactly two errors give rise to each syndrome
and the two errors corresponding to the measured syndrome have weights $n_a, n_b$,
then the ratio of amplitudes is $[\tan (\pi/2 - \alpha) ]^{(n_a - n_b)} \approx (\pi/2 - \alpha)(n_a - n_b)$,
and the SSRE is $\mathcal M_2 \approx (\pi/2 - \alpha)^2 (n_a - n_b)^2$ (see Supplementary Methods Section \ref{app:vanishing-rate-analytics}).
Fig.~\ref{fig:vanishing-rate}A shows the \ssre\, for classical simulations of circuits; the distribution is sharply peaked near this prediction.
Since the error weights $n_a, n_b$ controlling  the \ssre\, are drawn from a binomial distribution,
averaging over syndromes gives $\mathcal M_2 \propto N(\pi/2 - \alpha)^2 = f((\pi/2 - \alpha)\sqrt{N})$.
(See SM Section ~\ref{app:vanishing-rate-analytics} details.)
Fig.~\ref{fig:vanishing-rate}B  shows the syndrome- and circuit-averaged \ssre\, as a function of error angle $\alpha$;
Fig.~\ref{fig:vanishing-rate}C  shows the same quantity for experiments (see below).
Both show the predicted square-root scaling $\langle \mathcal M_2 \rangle = f((\pi/2 - \alpha) \sqrt N)$.
{
It is interesting to note that some of the same behavior occurs in the case of zero-rate surface codes \cite{Bravyi_coherent_2018}. In that case, the breakdown of the code and generation of magic in the logical qubit occurs at a threshold value below the Clifford point and can be understood through mappings to Anderson localization \cite{Venn2023}.
Here, we focus on random stabilizer codes for their conceptual simplicity and natural generalization to a finite-density of logical qubits, as considered in the next section.
}

\medskip 

\textit{Experiment:} We perform our experiments on IonQ's Aria quantum processor, made available through the QLab facility at the University of Maryland. We use 16 qubits for our experiments to limit the circuit depth and effects of noise. All quantum circuits, compiled into native gateset, were executed using API access. We provide further details on circuit execution in Supplementary Methods Section \ref{app:circuit-execution}. For the vanishing rate case, we run the encoding, error, and decoding circuit over $N$ physical qubits many times. Since we need to perform tomography on the single logical qubit, we append an appropriate basis change Pauli gate for each instance of random encoding circuit. Finally, we measure the entire register. 
Postselecting on syndrome outcomes is prohibitively expensive,
so we use the fact that the number of effective actions on the logical qubit (up to a global phase) is much smaller than the number of possible syndromes.
This allows us to group the syndrome into equivalence classes,
where elements in a class have the same effective action on the logical qubits.
These classes are identified by grouping the rotations using classical simulations. 
To mitigate incoherent errors, we project the density matrix of the logical qubit, obtained using tomography of a syndrome-class, to its maximum-eigenvalue eigenstate in post-processing. 
We then calculate the circuit- and syndrome-averaged magic
$\mathcal{M}(\alpha) = \langle |\overline{s}| \times \langle M_{C,\overline{s}} \rangle_{\overline{s} \in \mathcal{S}} \rangle_{C}$, where $\mathcal{S}$ denotes syndrome classes and $|\overline{s}|$ denotes the size of a syndrome class $\overline{s}$.

We present our experimental measurements in Fig.~\ref{fig:vanishing-rate}E for $N=8,12,$ and $16$, using $50$, $50$ and $30$ instances, respectively, of random circuits. The error-bars are obtained via bootstrap resampling (details in SM Section \ref{app:boostrapping-estimate}). We observe that, following the mitigation techniques discussed above, we can achieve a measurement of magic that qualitatively resembles the theoretical expectations.

\section{Constant Rate} 
\begin{figure*}
    \centering
    \includegraphics[width=\textwidth]{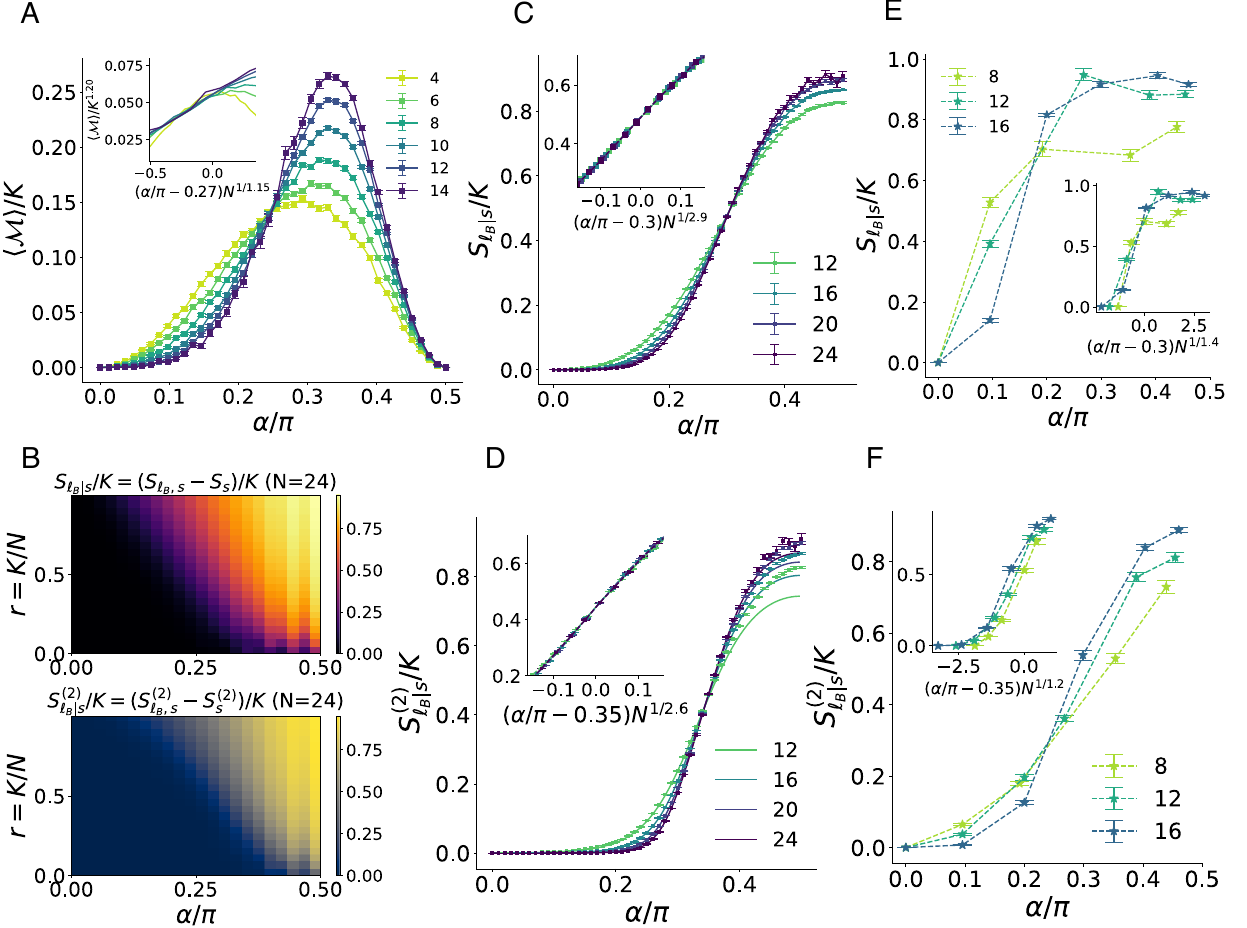}
    \caption{\textbf{Results for constant rate codes.} \textbf{A}: Density of magic (SSRE)  of the logical space and its scaling collapse (inset) plotted against the error rate $\alpha$, for code rate $r = K/N = 1/2$. The error bars are derived using standard error and are omitted in the scaling collapse (inset), where the x-axis is scaled as $(\alpha/\pi-\alpha_c/\pi)N^{1/\nu}$ with critical parameters $\alpha_c/\pi = 0.27(1)$ and $\nu = 1.15(4)$, and the y-axis is scaled as $\langle \mathcal{M}\rangle/K^{\gamma}$ with $\gamma = 1.20(8)$.  \textbf{B}: Phase diagrams of conditional entropy (upper) and its R\'enyi approximation (lower), without any basis minimization. \textbf{C}: Finite size scaling of the conditional entropy and its collapse (inset) computed numerically using simulations at $r=1/2$. The scaling collapse (inset) has critical parameters $\alpha_c/\pi = 0.304(2)$ and $\nu=2.9(2)$.  \textbf{\textbf{D}}: Finite size scaling of the R\'enyi approximation of the conditional entropy and its collapse (inset) computed numerically using simulations (displayed points) and analytics (solid line) at $r=1/2$.. The scaling collapse, computed with data from simulations, has critical parameters $\alpha_c/\pi = 0.347(1)$ and $\nu=2.6(2)$. \textbf{E}: Finite size scaling of the conditional entropy using data from experiments in IonQ Aria at $r=1/2$. The error bars are obtained using boostrap resampling. The scaling collapse (inset) uses critical exponents derived from numerical simulations of circuits with $d=N/2$, as shown in Fig.~\ref{fig:half-depth}. \textbf{F}: Finite size scaling of the R\'enyi-approximation of the conditional entropy and its collapse (inset) computed using experiments at $r=1/2$.}
    \label{fig:constant-rate-magic-full-panel}
\end{figure*}

At finite rate---that is, when the number of logical qubits $K$ scales as $K = rN$ with the number of physical qubits $N$---%
the finite-magic critical region displayed by the vanishing-rate code becomes an extended magical phase.
This magical phase is visible in Fig.~\ref{fig:overview}C, which shows the phase diagram of SSRE as a function of the code and error rate
in classical simulations for systems of $N \leq 14$ physical qubits.
Fig.~\ref{fig:constant-rate-magic-full-panel}A shows the density of SSRE at fixed rate $r = 1/2$ as a function of error rate $\alpha$,
again in classical simulations. We perform a free-parameter scaling collapse in the linear regime around the crossover, and find {\new{ $S/K^\gamma \propto (\alpha/\pi - \alpha_c/\pi)^{1/\nu}$ with $\alpha_c = 0.27(1), \nu = 1.15(4), \gamma = 1.20(8)$.}}
The precise values of the exponents are somewhat sensitive to the size of the regime in which we perform the collapse. {\new{ Moreover, free-parameter scaling suggests that the magic scales at the critical point as $N^{1.2}$. But physical intuition suggests that, since the scaling dimension is zero below the critical point and one above the critical point, scaling dimension at critical point should be within $[0, 1]$. In SM Section \ref{sec:constrained-gamma}, we present Fig.~\ref{fig:constant-rate-magic-full-panel}A but with $\gamma$ constrained to $[0,1]$.}}

The scaling collapse indicates that the transition from non-magical to magical is indeed a phase transition, not a crossover.

Since the SSRE is an expensive quantity to measure for finite-rate codes, we use the conditional entropy of the logical state as a diagnostic for the phases. In Fig.~\ref{fig:constant-rate-magic-full-panel}B (upper), we show the phase diagram for conditional entropy density as a function of code rate and error-rate. The conditional entropy, without any minimization over basis, serves as an upper-bound to the basis-minimized conditional entropy, which is a genuine measure of magic.
{
Moreover, for small $\alpha$ we have a priori reason to believe that the computational basis is the minimum-entropy basis:
the logical state begins in a computational basis state,
and when the $\alpha$ is small it is weakly perturbed,
so it remains closer to the computational basis state than any other stabilizer state.
This is no longer true at the Clifford point $\alpha = \pi/2$.
There the minimum-entropy basis is no longer the computational basis, but some other stabilizer basis;
in that basis the measurement entropy is 0,
but in the computational basis the measurement entropy is extensive.
We discuss the relationship between measurement entropy, error correction, and decoder breakdown in Supp.~\ref{supp:decoder-breakdown}.
}

In Fig.~\ref{fig:constant-rate-magic-full-panel}C, we present finite rate scaling, obtained through simulations, of the conditional entropy at code rate $r=1/2$. For each datapoint in numeric simulations of sizes $N= 12, 16, 20$ and $24$, we simulate $5000, 5000, 500$ and $50$ circuits, respectively. The procedure used to extract the critical parameters and their errors is described in Supplementary Method Section \ref{app:finite-size-scaling}. We observe that this critical error rate $\alpha_c$ and critical exponent $\nu$ both differ from the SSRE. {\old{\color{red} Such a difference in critical point and scaling exponent is a common feature of phase transitions in monitored quantum circuits; it comes about because different quantities correspond to different statistical mechanical models \cite{jian_measurement-induced_2020, bao2020theory} and because different measures quantify resources differently. Additionally, while the scaling collapses of Fig.~\ref{fig:constant-rate-magic-full-panel} are the best free-parameter collapses, we cannot rule out other collapses. The goal of our work is not to unify these various diagnostics of magic or to resolve different detailed phase transitions, but rather to present the existence of a transition in the non-stabilizerness or magic.}}
  {\new{ Given the limited understanding of this model's phases, compounded by the challenges with numerical studies of large systems, it is difficult to have confident estimates of the critical point and exponent. In particular, we cannot conclude if the transition point coincides (or not) with that of the SSRE, as this depends drastically on the choice of scaling hypothesis. This discussion is left as an outlook. Indeed, our work aims not to unify these diagnostics or resolve their detailed phase transitions but rather to present a transition on non-stabilizerness or magic in the first place.}}

\textit{Analytical calculations:}
We also observe that the R\'enyi conditional measurement entropy, the R\'enyi analogue of the Shannon conditional measurement entropy, exhibits similar phases, as shown in Fig.~\ref{fig:constant-rate-magic-full-panel}B (lower). The circuit-averaged conditional R\'enyi entropy is defined as 
\begin{equation}
\mathbb{E}_C \left[ -\log \sum_{x \in \{0,1\}^N} \\ \left(p^{\alpha,C}_{(\ell,s)}(x)\right)^2 + \log \sum_{x \in \{0,1\}^{N-K}} \left(p^{\alpha,C}_{s}(x)\right)^2  \right],
\nonumber
\end{equation}
where $p_{s}^{\alpha,C}(x)$ denotes the probability of measuring $x$ in the syndrome register $s$, for a state produced with a Clifford encoder $C$ and coherent rotations of strength $\alpha$. Similarly the $(\ell,s)$ subscript denotes measurement in the joint syndrome and logical register. For large systems, we can use a typicality argument to assume that the circuit-to-circuit variation in distribution of measurement outcomes is negligible; this allows us to interchange logarithm with expectation over $C$ in the equation above. Finally, we can calculate $\mathbb{E}_C \sum_x (p^{\alpha,C}(x))^2 $ using Clifford averaging via Schur-Weyl duality (see  SM Section \ref{app:renyi-analogue-analytics} for details). 

Fig.~\ref{fig:constant-rate-magic-full-panel}D shows the finite-size scaling for \rcme\, for system sizes $N \le 24$ comparable to those used in classical simulation and experiment, at code rate $r=1/2$. The analytical result is plotted as solid lines. The datapoints represent circuit averages computed via numerical simulations. For each datapoint in numeric simulations of sizes $N= 12, 16, 20$ and $24$, we simulate $5000, 5000, 500$ and $50$ circuits, respectively. Indeed, the analytical computation matches the exact numerics for large systems for which the typicality assumption is true. While our analytics lets us access arbitrarily large sizes for this observable, we perform the scaling collapse over sizes that are experimentally accessible. 
The procedure used to perform the scaling collapse and extract related errors is described in Supplementary Method Section \ref{app:finite-size-scaling}.

\textit{Experiment:}
The random Clifford code was implemented on up to 16 qubits of IonQ's 32-qubit Aria device.
(The implementation was limited not by the number of qubits available but by circuit depth,
which in turn was limited by gate noise.) 
The \ssre\, is not accessible in experiments on finite-rate codes,
because it requires full state tomography.
The conditional measurement entropy, by contrast,
requires only computational basis measurements.
These computational basis measurements undergo postprocessing similar to linear cross-entropy benchmarking in random circuit sampling experiments\cite{arute2019quantum},
using information from classical simulation.
The reported entropies are $S_X = -\sum_x p(x) \log \tilde{p}(x)$, where $\tilde{p}(x)$ is the probability anticipated from classical simulation, and $p(x)$ are experimentally obtained distributions projected onto the support of the ideal distribution $\tilde{p}(x)$. 

Figures \ref{fig:constant-rate-magic-full-panel}E and \ref{fig:constant-rate-magic-full-panel}F show the resulting Shannon and R\'enyi entropies, respectively. As expected, the scaling of the entropies with respect to system size is inverted across the threshold. Recall that the experiments were performed using circuits of depth $d=N/2$. The scaling collapse (insets) in the experimental data use the critical parameters derived from numerical simulations for $d=N/2$ circuits, as shown in Fig.~\ref{fig:half-depth}B,C in SM Section \ref{app:half-depth-circuits}. For each experimental datapoint of sizes $N=8,12$ and $16$, we execute $50,50$ and $20$ different circuit instances, respectively. The errorbars were estimated using boostrap resampling, discussed in SM Section \ref{app:boostrapping-estimate}.

\section{Discussion}

We have observed that a random Clifford code subject to coherent errors displays a phase transition in magic. Concentrating magic of a large system to a smaller subsystem can be difficult, as has also been shown in \cite{leonePhaseTransitionStabilizer2023}. In our model, measuring the syndromes of a random Clifford code concentrates magic in the logical space if the error rate or the code rate is above a critical value,
and suppresses magic below the threshold. This result establishes a connection between the resource theory of stabilizer computation, i.e., magic,
and the study of decoder breakdown in quantum error correction codes
via the basis-minimized measurement entropy.

In this work, we study phases of magic for small systems for which calculating measures of magic is tractable. In general, non-stabilizerness is difficult to measure. Measures of magic usually require exponentially many measurement samples and often need extensive classical processing, making them intractable for large systems. Our work, however, suggests the possibility of diagnostics, like the conditional entropy, that can be estimated efficiently using a small number of samples and classical post-processing. In the future, such measures can be used to study the phase transition in larger systems to better approximate the thermodynamic limit. 

Phase transitions in magic---both ours and the theoretical predictions of \cite{leonePhaseTransitionStabilizer2023}---indicate that existing measurement-induced phase transitions sit in a broader landscape of information theoretic phase transitions.
In each case, the phase transition arises from the competition between three channels---a channel that generates the resource (whether entanglement or magic), a channel that generates correlations, and a channel that destroys the resource---that fail to commute.
In the phase transition shown here, the correlation-generating channels are the encoding Clifford operations, the resource-generating channels are the rotations $R_z(\alpha)$ of the noise layer, and the resource-destroying maps are syndrome measurements.
In the phase transition of \cite{leonePhaseTransitionStabilizer2023} the correlation-generating channels are layers of random Clifford gates, the resource-generating channels are interspersed T gates, and the resource destroying maps are partial traces.
In the measurement-only entanglement phase transition of \cite{ippolitiEntanglementPhaseTransitions2021}, all channels are projective measurements:
nonlocal projective measurements generate entanglement as well as correlation, while onsite measurements destroy the resource.
We conjecture that any information-theoretic setting with this structure of three competing channels can be made to show a phase transition.  
{ A related question concerns the nature of universality in these information theoretic transitions at their critical points.  It is currently unclear whether the magic phase transitions studied are indeed part of a universality class of critical phenomena.}


Our result also suggests that error correction together with sufficiently well-characterized coherent noise can create useful magic states.
In the magical phase the syndrome measurements move magic from the physical qubits, where non-Clifford gates like single-qubit rotations are easy, 
to the logical qubits, where non-Clifford gates are typically hard.
Syndrome-dependent Clifford unitaries may then transform these states into states suitable as inputs to existing magic state distillation protocols. In this case, an outstanding challenge is the decoding problem of identifying the right Clifford unitary given a code and a syndrome. 
Notably, such unitaries are efficiently computable under a wide-range of circumstances for zero-rate topological surface codes \cite{Bravyi_coherent_2018}.
If this can be done more generally, the magical phase may improve the efficiency of magic state distillation, thereby reducing overhead in quantum computation algorithms where magic state distillation is the bottleneck.

\section*{Acknowledgments}
We thank Ken Wright, Melanie Hiles, and Jason Nguyen from IonQ for their assistance, and Liudmila Zhukas, Nicole Yunger Halpern, William Braasch, and Brayden Ware for comments on the manuscript;
{ we are also grateful to an anonymous reviewer for a thorough, careful, and thoughtful review}.
This work is supported by the NSF STAQ program; the NSF QLCI RQS Program OMA-2120757; the DOE QSA program DE-FOA-0002253; and the AFOSR MURIs on Dissipation Engineering in Open Quantum Systems, Quantum Measurement/Verification, and Quantum Interactive Protocols.
CDW thanks DOE-ASCR Quantum Computing Application Teams program for support under fieldwork proposal number ERKJ347.
The experiments were performed on the IonQ Aria system through the UMD/IonQ Q-Lab consortium. 

\section*{Data Availability}
{ All the data used in this work can be found in a Zenodo repository \href{https://zenodo.org/doi/10.5281/zenodo.7847794}{DOI: 10.5281/zenodo.7847794}}.

\section*{Disclaimer}
Certain commercial equipment, instruments, or materials are identified in this paper in order to specify the experimental procedure adequately. Such identification is not intended to imply recommendation or endorsement by the National Institute of Standards and Technology, nor is it intended to imply that the materials or equipment identified are necessarily the best available for the purpose.

\bibliographystyle{science}
\bibliography{citations}


\clearpage
\newpage 
\onecolumngrid
\renewcommand{\thefigure}{S\arabic{figure}}
\setcounter{figure}{0}
\renewcommand{\thesection}{\Alph{section}}
\setcounter{section}{0}
\part*{Supplementary Material}
\section{Basis-Minimized measurement entropy as a measure of magic}
\label{app:basis-minimized-entropy}
Here we show that the basis-minimized conditional entropy is a good measure of non-stabilizerness. Consider a pure state $\ket{\psi}$.
Measuring this state in the computational basis produces a classical bitstring $x$ drawn from the Born probability distribution $p(x) = |\braket{x|\psi}|^2$. We can instead choose to measure in a stabilizer basis different than the computational basis, by rotating the state using a Clifford unitary. The basis-minimized measurement entropy is the entropy of this probability distribution, minimized over bases:
\begin{equation}
    z^*(\psi) = \min_{C}\left[ -\sum_x |\braket{x|C^\dagger|\psi}|^2 \log |\braket{x|C^\dagger|\psi}|^2 \right]
    \label{eq:bme}
\end{equation}
We would like to show that the the basis-minimized conditional entropy is i) zero for stabilizer states, ii) non-increasing under Clifford unitaries and iii) sub-additive. 
\begin{enumerate}
    \item \textbf{Faithfulness}: If $\ket{\psi}$ is a pure stabilizer state, there exists some $C^*$ such that $\ket{\psi}  = C^*\ket{0}$. We can therefore choose $C = (C^*)^\dagger$ in \eqref{eq:bme} to get $z^*=0$. 

    \item {\textbf{Stability under Clifford Unitaries}: Applying anther Clifford gate $C'$ to some $\ket{\psi}$ should not change $z^*$. Suppose the Shannon entropy of $\ket{\psi}$ is minimized for some $C^*$. Now the Clifford operation $C'$ takes the state to $\ket{\psi'} = C'\ket{\psi}$. The quantity $z^*$ is now
\begin{align}
\begin{split}
    z^*(\psi') &= \min_{C}\left[ -\sum_x |\braket{x|\psi'}|^2 \log |\braket{x|\psi'}|^2 \right] \\
    &= \min_{C}\left[ -\sum_x |\braket{x|C'|\psi}|^2 \log |\braket{x|(C')|\psi}|^2 \right] \\
\end{split}
\end{align}
We can recover the original $z^*(\psi)$ by taking $C = C' C^*$. Therefore $z^*$ does not increase under Clifford gates.}

    \item {\textbf{Subadditivity}:} Given a product state $\ket{\psi} = \ket{\phi}\otimes\ket{\sigma}$, we have
    \begin{align}
    \begin{split}
          z^*(\ket{\psi}) &= \min_{C}\left[ -\sum_x |\braket{x|C|\phi, \sigma}|^2 \log |\braket{x|C|\phi, \sigma}|^2 \right] \\ &\leq \min_{C_1\otimes C_2}\left[ -\sum_x |\braket{x|C_1\otimes C_2|\phi, \sigma}|^2 \log |\braket{x|C_1\otimes C_2|\phi, \sigma}|^2 \right] \\&= \min_{C_1\otimes C_2}\left[ -\sum_{x_1,x_2} |\braket{x_1,x_2|C_1\otimes C_2|\phi, \sigma}|^2 \log |\braket{x_1,x_2|C_1\otimes C_2|\phi, \sigma}|^2 \right] 
          \\&= \min_{C_1}\left[ -\sum_{x_1} |\braket{x_1|C_1|\phi}|^2 \log |\braket{x_1|C_1|\phi}|^2 \right] \\ & \quad  +  \min_{ C_2}\left[ -\sum_{x_2} |\braket{x_2|C_2|\sigma}|^2 \log |\braket{x_2|C_2|\sigma}|^2 \right] \\
          &= z^*(\ket{\phi}) + z^*(\ket{\sigma})
      \end{split}
    \end{align}
    Here, in the second step we confine the minimization to Clifford unitaries of the form $C_1\otimes C_2$. In the fourth step, we use the independence of the probability distribution across the two halves of the quantum state to decompose the Shannon entropy. 
\end{enumerate}

It is helpful to prove the converse of faithfulness, i.e. that $z_*(\psi) = 0$ implies that $\ket \psi$ is a stabilizer state.
To see this, note that if $z_*(\psi) = 0$, then there is some Clifford $C$ such that all the overlaps $\braket{x|C^\dagger|\psi}$ are 0 or 1.
This in turn means that $C^\dagger \ket \psi$ is one of the computational basis states, call it $\ket {x_*} = C^\dagger\ket{\psi}$,
and
\begin{align}
    \ket \psi = C \ket{x_*}
\end{align}
is itself a stabilizer.


\section{Distribution of Syndromes in Vanishing Rate and Finite Rate Codes}
\label{app:syndrome-distribution}
\begin{figure}[!ht]
    \centering
    \includegraphics[width=0.6\textwidth]{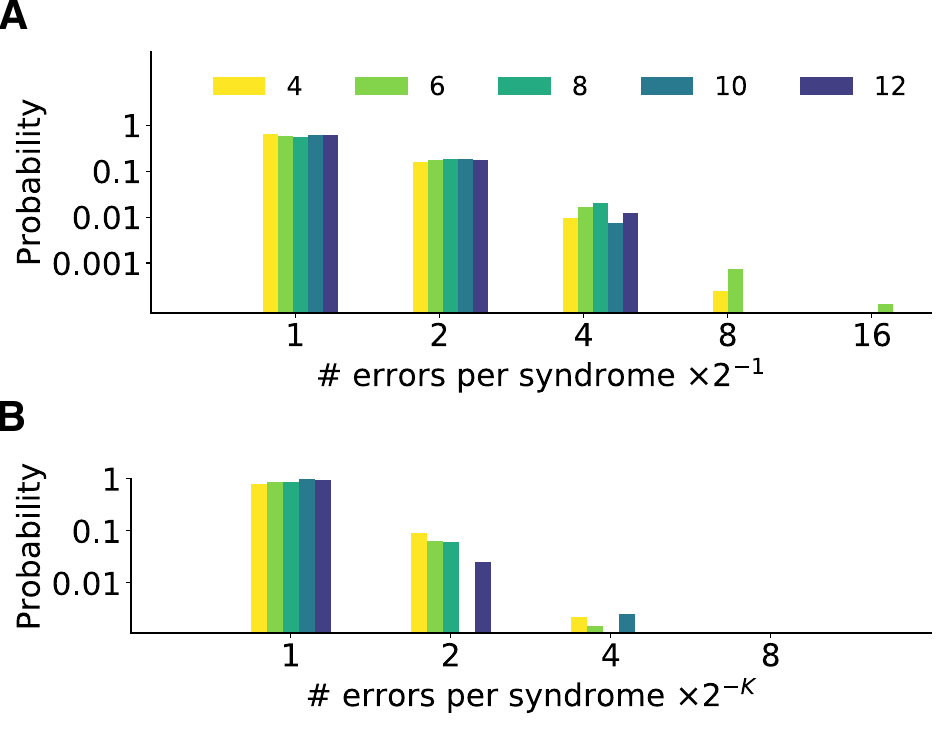}
    \caption{\textbf{Distribution of Errors}  The distribution of error per syndrome for vanishing rate codes (\textbf{A}) and constant rate codes $(\textbf{B})$. We observe that in both cases, with high probability, the errors are uniformly distributed across syndromes, such that there are $2^{K}$ unique errors per syndrome. Colorbars give the number of qubits $N$. }
    \label{fig:distribution-of-errors}
\end{figure}

In Fig.~\ref{fig:distribution-of-errors}, we plot the distribution of errors per syndrome for the random codes used in our model. The data for the plots is generated using the following procedure: for each circuit, we implement all $2^N$ Pauli errors. Each error is a Pauli string of identity and $Z$ operator, eg $IIZZII$. For each error, we record what syndromes we measure at the end of the circuit. Having recorded the syndromes for all Pauli errors, we then count the number of errors that result in a particular syndrome.

We observe that, with high probability, the $2^N$ unique errors are uniformly distributed across the $2^{N-K}$ syndromes. In our analysis of vanishing rate codes, we assume that the codes have exactly two errors per syndrome. 

\section{Basis-minimized measurement entropy and decoder breakdown}\label{supp:decoder-breakdown}

The basis-minimized measurement entropy is a direct probe of the breakdown of the optimal Clifford decoder.
To see why, imagine storing a classical bit string as a computational basis state,
encoding it using the Clifford circuit of our model,
subjecting it to error,
performing the conjugate of the the encoding circuit and error measurements,
and attempting to recover it by measuring the logical state in the computational basis.
 Without error, there will be exactly one possible syndrome and one possible logical measurement outcome.
With error, there may be many:
approximately
\begin{align}
Z_{\bm s,\text{comp.} } = \exp S[p_{\bm s, \text{comp}}]
\end{align}
possible bitstrings correspond to syndrome $\bm s$,
where $p_{\bm s,\text{comp}}$ is the Born probability distribution of outcomes of measuring the logical state corresponding to syndrome $\bm s$ in the computational basis,
and $S$ is the entropy.
If $Z_{\bm s, \text{comp}}$ is not small, measuring in the computational basis may produce any of number of outcomes---not just the initial stored bitstring.

If the noise is coherent, the net effect of the channel is to apply a unitary or weak projection to the logical state.
In this case measuring in a different basis may recover the information;
the basis change is known as a decoder.
The best Clifford decoder gives
\begin{align}
    Z_{\bm s} = \min_B \exp S[p_{\bm s, B}]
\end{align}
possible logical measurement outcomes,
where now $p_{\bm s, B}$ is the Born probability distribution of outcomes of measuring the logical state corresponding to syndrome $\bm s$ in the basis given by Clifford circuit $B$.
Across syndromes, the typical number of possible logical measurements is given by the average over syndromes
\begin{align}
\begin{split}
     Z_{l,\text{typical}} &= \exp \left\langle\ln Z_{l;\bm s}\right\rangle_{\bm s} \\
     &= \exp \left\langle \min_B S[p_{\bm s,B}] \right\rangle_{\bm s}
 \end{split}
\end{align}
When this quantity is materially greater than 1, i.e.
\begin{align}
\label{eq:breakdown-s-dep}
\varepsilon < \left\langle\min_B S[p_{\bm s,B}]\right\rangle_{\bm s}
\end{align}
for some small $\varepsilon$,
the code has broken down and cannot even store classical information: it has probability $\sim e^{\varepsilon}$ of irrecoverably muddling the input bitstring.

Storing classical information in this way is a weaker condition than storing quantum information.
Moreover, while a large measurement entropy indicates that classical information is irrecoverable, small measurement entropy does not indicate that classical information is recoverable, because syndrome measurement may project the logical state to a pure stabilizer state different from  the initial state.

The condition \eqref{eq:breakdown-s-dep} implicitly allows the choice of measurement basis for the logical space to vary with syndrome. An intelligent decoder will use the syndrome to pick an optimal basis. A more primitive decoder will pick a fixed basis, and use that for all syndromes. Such a decoder will fail if
\begin{align}
\varepsilon < \min_B \left\langle S[p_{\bm s,B}]\right\rangle_{\bm s}\;.
\end{align}
The simplest decoder of all leaves the logical state in the computational basis; it fails if 
\begin{align}
\varepsilon < \left\langle S[p_{\bm s,B}] \right \rangle_{\bm s} \;.
\end{align}
This is the quantity we treat in the main text.

\section{Analytical Estimate of Magic in the Vanishing Rate Code}
\label{app:vanishing-rate-analytics}
In this section we estimate the magic our model at vanishing rate---that is, for a single logical qubit.
We first compute the action of the channel on that logical qubit;
we find that near the Clifford point, $\alpha = \pi / 2$,
it is a unitary with probability $1/2$.
(Whether or not it is a unitary depends on circuit and syndrome measurement outcome.)
We then pass a stabilizer state through the channel and compute the magic of the result.
For $\epsilon = \pi/2 - \alpha$
we find that the magic resulting from a single circuit (C) and syndrome measurement outcome is
\begin{align}
    M_{2;C} = (n\epsilon)^2\;,
\end{align}
where the integer $n$ is determined by the weights of the errors corresponding to the measurement outcome.
Averaging across measurement outcomes this becomes
\begin{align}
    \expct{M_2} = \frac 1 4 N \epsilon^2
\end{align}
for $N\epsilon^2 \ll 1$, and $\mathbb{E}({M_2}) = f(N\epsilon^2)$ in general.

\subsection{Action of the channel on the logical space}\label{ss:vanishing-unitary}
\subsubsection{The Clifford point}
Consider the vanishing-rate (single qubit) code at the Clifford point $\alpha = \pi/2$. Let the Clifford encoder used in the circuit be $C$. The error unitaries can be expanded as
\begin{align}
  U_\alpha &= C^\dagger \prod_{j=1}^N e^{i\alpha\sigma_z/2} C  = \prod_j [\cos \alpha /2 +i \sin \alpha /2\; \tilde{\sigma}_z^{(j)}] = 2^{-N/2}\sum_{\bm a \in \mathbb{B}^N} i^{n_{\bm a}}\tilde{\sigma}^{\bm{a}},
\end{align}
where $\tilde{\sigma}_z^{(j)} = C^\dagger {\sigma}_z^{(j)}C$ and $\bm a$ are length-$N$ bitstrings, $\sigma^{\bm{a}}$ is
\begin{align}
  \tilde{\sigma}^{\bm{a}} = \prod_{j = 1}^N \left(\tilde{\sigma}_z^{(j)}\right)^{a_j}\qquad\text{ with 
 } \quad \left(\tilde{\sigma}_z^{(j)}\right)^0 = 1
\end{align}
and $n_{\bm a}$ counts the number of error Paulis in $\err a$
\begin{align}
  n_{\bm a} = \sum_j a_j\;.
\end{align}
Below we rewrite $\tilde{\sigma}$ operators as simply $\sigma_z$, noting that these are now highly non-local operations due to conjugation by $C$.

There are $2^N$ such bitstrings, hence $2^N$ such errors
(including the trivial ``error'', the identity operator $\bm a = 0$). When we measure all the $N-1$ syndrome qubits, we see one of $2^{N-1}$ syndromes.
(Since vanishing rate codes have $2$ errors per syndrome with high probability, we observe inFig.~\ref{fig:distribution-of-errors}, we restrict this analysis to to codes in which each syndrome corresponds to exactly $2$ errors.
In numerics this can be done by postselection.
)
Say the measured syndrome is $s$,
so projective measurement onto that syndrome is $P_s$. Let us also denote the two errors giving rise to $s$ by $\sigma_z^{\bm{a}}$ and $\sigma_z^{\bm{b}}$. If $\ket{\psi}$ is the state after the encoding Clifford unitary, noise layer and the conjgate of the Clifford encoder, the state after the syndrome measurement is then proportional to
\begin{align}
 P_s\Big(\err a + i^{n_{\bm b} - n_{\bm a}} \err b\Big)\ket \psi.
\end{align}

The errors $\sigma^{\bm a,\bm b}$ are Pauli strings.
Let $\sigma ^{\bm a, \bm b}_1$ be the Pauli in each Pauli string that acts on site 1 in the code basis. 
Then, the effective action of the error channel and syndrome measurement on the logical qubit 1 is
\begin{align}
  \label{eq:logical-unitary-clifford}
  K = \err{a}_1 + i^{m}\err{b}_1\;,
  \quad m &= \zeta(\bm b) - \zeta(\bm a) + n_{\bm b} - n_{\bm a} \;,
\end{align}
where $\zeta(a)$ and $\zeta(b)$ account for the fact that $\sigma _{\bm a}$ and $\sigma _{\bm b}$ can put different phases on the states corresponding to our syndrome $s$.%
\footnote{
  Consider for example $N = 2$ with (non-commuting) errors $\sigma _{\bm a} = xx$ and $\sigma _{\bm b} = xy$.
  These give the same syndrome with different phases;
  the $\zeta $ term encodes those phases.
}
This action $K$ is (proportional to a) unitary if $m$ is odd and $[\err a, \err b] = 0$,
or $m$ is even and $\{\err a, \err b\} \ne 0$.
Otherwise $K$ is a projector.
We expect this will happen with probability $1/2$.
If we average over syndromes, the result is a channel
\begin{equation}
    \mathcal E_{k = 1}(\rho) = \frac 1 2 P \rho P + \frac 1 2 U\rho U^\dagger
\end{equation}
where $P$ is a projector onto some stabilizer state and $U$ is a unitary deducible from the syndrome
and the coding circuit $C$.

\subsubsection{Away from the Clifford point}
Now move slightly away from the clifford point---take $\alpha \ne \pi/2$.
The action on the logical space \eqref{eq:logical-unitary-clifford} becomes
\begin{equation}
    K \propto \left[{\err a}_1 + i^{m} (\tan \alpha/2)^{n_{\bm b} - n_{\bm a}} {\err b}_1\right]\;.
\end{equation}
Once again this is unitary for
$m$ odd and $[\err a, \err b] = 0$,
or $m$ even and $\{\err a, \err b\} \ne 0$,
and once again these case arises with probability $p \approx \frac 1 2$.
In the other cases, $K$ is (unitarily equivalent to) a weak projection 
\begin{align}
    K = P' &\sim \begin{bmatrix} 1 & 0 \\ 0 & \frac 1 2 \left[1 - (\tan \alpha/2)^{n_{\bm b} - n_{\bm a}} \right]\end{bmatrix} 
    \approx
    \begin{bmatrix} 1 & 0 \\ 0 & \frac {n_{\bm b} - n_{\bm a}} 2 \epsilon \end{bmatrix}\;.
\end{align}

\subsection{Magic}

To see how far away from the set of stabilizer states the error and measurement take us,
let us apply $K$ to an initial stabilizer state $\ket 0$ on the logical qubit.
(We can check $\ket 0$ without loss of generality, because other initial stabilizer states correspond to different elements of the ensemble of encoding circuits.)
If $K$ acts as a weak projector, the result is again a stabilizer state.

But now suppose $K$ acts as a unitary.
If $[\err a, \err b] = 0$ then (up to a Clifford operator)
\begin{equation}
K \ket 0 = \frac 1 {\sqrt{2} } (1 + i^m) \ket 0\;:
\end{equation} 
(with $m$ odd): $K$ maps $\ket 0$ to another stabilizer state.
If on the other hand $\{\err a, \err b\} = 0$, then up to a Clifford unitary
\begin{equation}
K \ket 0 = \frac 1 {\sqrt{2}} \Big[\ket 0 + i^m (\tan \alpha/2)^{n_{\bm b} - n_{\bm a}} \ket 1 \Big]\;.
\end{equation} 
For $\epsilon = \pi / 2- \alpha \ll 1$, taking $m = 0$, this is
\begin{equation}
\label{eq:vanishing-K-taylor}
K \ket 0 \approx \frac 1 {\sqrt 2} e^{i(n_{\bm b} - n_{\bm a})\epsilon \sigma_x}\Big[ \ket 0 + \ket 1 \Big]\;.
\end{equation}
(If $m \ne 0$ then $\sigma^x$ becomes $-\sigma^x$ or $\pm \sigma^y$.)
This is an $(n_{\bm b} - n_{\bm a}) \epsilon$ rotation away from the stabilizer state $\ket 0 + \ket 1$.
The second R\'enyi entropy of magic is 
\begin{align}
  M_2 = [n_{\bm b} - n_{\bm a} ]^2 \epsilon^2\;;
\end{align}
Fig.~\ref{fig:vanishing-rate}A shows the magic for individual syndromes, together with this prediction.
The average over measurement outcomes and circuits is therefore
\begin{align}
\begin{split}
    \expct{M_2} &= p_{\text{unitary}} \times \frac 1 2 \epsilon^2 \expct{(n_{\bm b} - n_{\bm a})^2} \\
    &= \frac 1 4 N \epsilon^2\;.
\end{split}
\end{align}

To see the second line, note that the distribution from which $n_{\bm a, \bm b}$ are drawn is close to the binomial $p(\bm n) = {N \choose n}$.
(Corrections to this distribution are higher-order in $\epsilon$.)
Consequently $n_{\bm a, \bm b}$ have mean $N/2$ and variance $N/4$,
so---assuming independence---$(n_{\bm b} - n_{\bm a})$ has mean 0 and variance $\frac 1 2 N$.

For $(n_{\bm b} - n_{\bm a})\epsilon \not\ll 1$ the Taylor series approximation of \ref{eq:vanishing-K-taylor} still breaks down,
but the characteristic scale is still
\begin{equation}
    (n_{\bm b} - n_{\bm a})\epsilon \sim \sqrt{N} \epsilon\;,
\end{equation}
so one expects a scaling collapse when we plot $M_2$ against $\epsilon \sqrt N$.
Fig.~\ref{fig:vanishing-rate}B shows this scaling collapse.

\section{Numerics for $d=N/2$}
\label{app:half-depth-circuits}
\begin{figure}
    \centering
    \includegraphics[width=\textwidth]{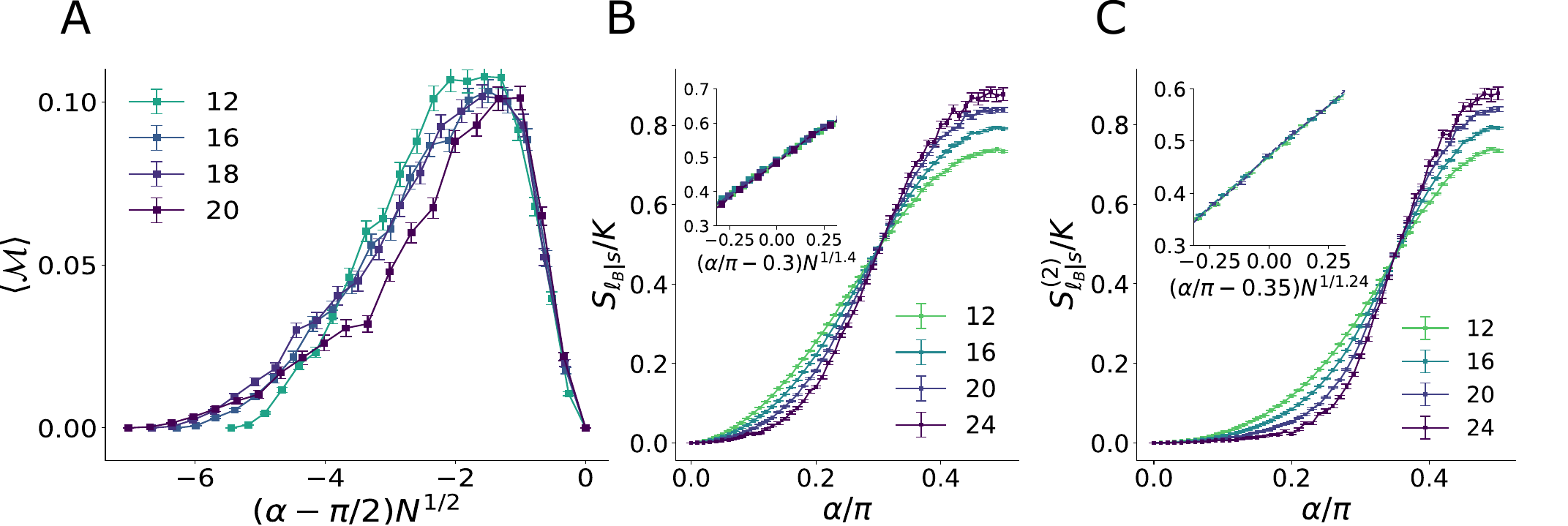}
    \caption{Numerical Simulations for circuits with depth $d=N/2$. \textbf{A}: \ssre for vanishing-rate code. Like with $d=N$ circuits, this exhibits a $\sqrt{N}$ scaling near the critical point at $\alpha=\pi/2$. \textbf{B}: Finite size scaling of the conditional entropy and its collapse (inset) computed numerically using simulations at code rate $r=1/2$. The error bars are omitted in the scaling collapse (inset) which has critical parameters $\alpha_c/\pi = 0.300(2)$ and  $\nu=1.40(6)$. These critical parameters are used in the collapse of experimental data in Fig.~\ref{fig:constant-rate-magic-full-panel}{E}. \textbf{C}: Finite size scaling of the \Renyi-approximation of the conditional entropy and its collapse (inset) computed numerically using simulations at code rate $r=1/2$. The error bars are omitted in the scaling collapse (inset) which has critical parameters $\alpha_c/\pi = 0.351(1)$ and $\nu=1.24(4)$. These critical parameters are used in the collapse of experimental data in Fig.~\ref{fig:constant-rate-magic-full-panel}{F}.}
    \label{fig:half-depth}
\end{figure}

While the numerical data presented in the main text (Fig.~\ref{fig:overview}C , Fig.~\ref{fig:vanishing-rate}AB and Fig.~\ref{fig:constant-rate-magic-full-panel}ABCE) used simulations of circuits of depth $d=N$, the experiments were performed with circuits of depth $d=N/2$ to reduce the effects of noise. In Fig.~\ref{fig:half-depth}, we present numerics using simulations with circuits of depth $d=N/2$ for quantities we experimentally probe, namely magic in vanishing rate codes and conditional entropies in constant rate codes. The critical exponents so obtained are used for scaling collapse for the experimental data presented in the maintext. 

\begin{figure}
    \centering
    \includegraphics[width=\textwidth]{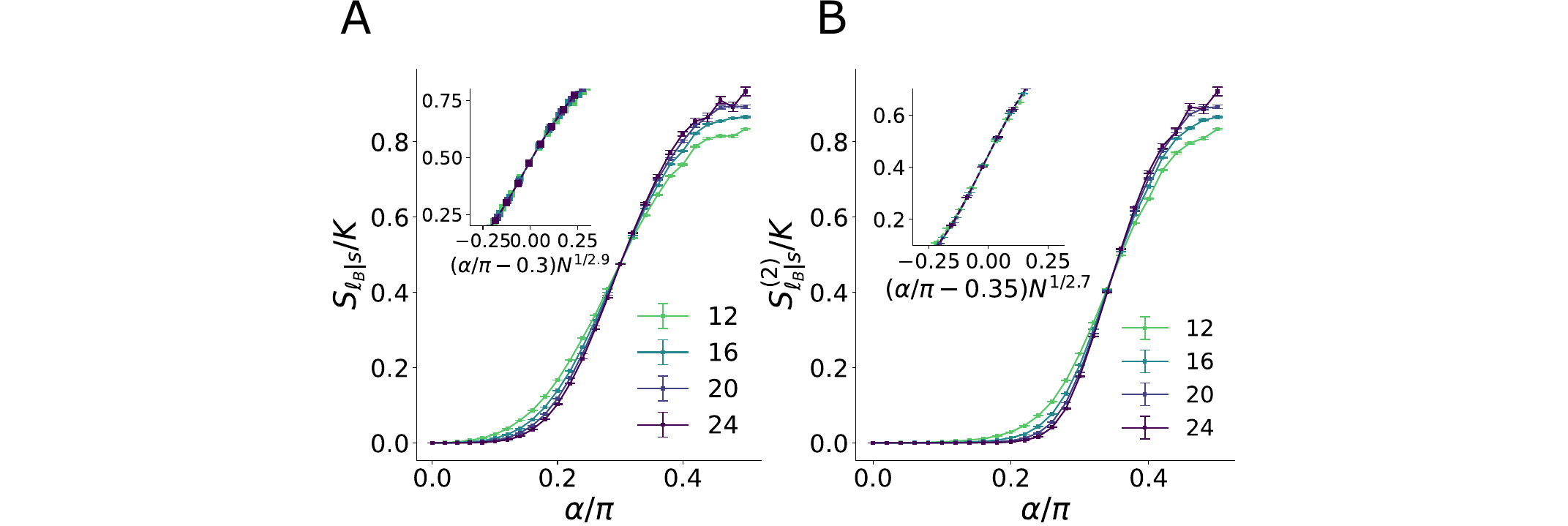}
    \caption{Numerical Simulations for circuits with depth $d=2N$.  \textbf{A}: Finite size scaling of the conditional entropy and its collapse (inset) computed numerically using simulations at code rate $r=1/2$. The scaling collapse (inset) has critical parameters $\alpha_c/\pi = 0.302(3)$ and $\nu=2.9(2)$.  \textbf{B}: Finite size scaling of the \Renyi-approximation of the conditional entropy and its collapse (inset) computed numerically using simulations at code rate $r=1/2$. The scaling collapse (inset) has critical parameters $\alpha_c/\pi = 0.348(3)$ and $\nu=2.8(4)$.}
    \label{fig:revised-double-depth}
\end{figure}

Note that the critical exponents of conditional entropies in Fig.~\ref{fig:half-depth}(B,C) are different than the critical exponents for circuits with $d=N$. We expect the critical exponents to converge for sufficiently deep encoding circuits -- that is once the circuits start forming good error correcting codes. In  Fig.~\ref{fig:revised-double-depth}, we present numerics on $d=2N$ circuits, which results in critical exponents close to $d=N$ circuits. This suggests that the $d=N$ circuits can adequately capture ensemble-average properties. 

\section{Details on Circuit Execution}
\label{app:circuit-execution}
The circuits were produced by randomly sampling Clifford encoders. Each encoder has depth $d$, where a unit of depth consists of a layer of $N$ single-qubit gates and a layer of $N/2$ disjoint pairs of entangling gates. The single qubit gates are sampled from the set of 24 single-qubit Clifford gates. The entangling gate is chosen to be the fixed-angle M\o{}lmer-S\o{}rensen gate, $MS(\pi/2)$. After selecting the gate sequence for each circuit, the encoder and decoder are optimized separately. After optimization, the circuits are compiled natively to a gateset comprising GPi, GPi2 and MS gates, as described in IonQ Documentation \cite{IonQDocumentation}. As a part of execution, the circuits were further augmented with single-qubit gates to minimize noise, using a firmware-level protocol described in \cite{maksymov2023enhancing}.

\section{Finite Size Scaling}
\label{app:finite-size-scaling}
To obtain the critical parameters of the scaling collapse, we assume that the quality of interest $f(\alpha, N)$ is a function of error rate $\alpha$ and the code size $N$ and can be expanded as 
\begin{equation} 
f(\alpha,N) \approx A + Bx + Cx^2 \qquad x = (
\alpha-\alpha_c)N^{1/\nu}
\end{equation}
Using data collected using numerical simulations $y_{\alpha,N}$, we minimize the following mean squared error to obtain the estimate for the critical parameters $\alpha_c$ and $\nu$:
\begin{equation}
  \hat{\alpha}_c, \hat{\nu} = \argmin_{\alpha_c, \nu} \min_{A, B, C} \sum_{y_{\alpha,N}}(A + Bx + Cx^2  - y_{\alpha,N})^2 \qquad \text{ with  } x = (
\alpha-\alpha_c)N^{1/\nu}
\end{equation}
To obtain the error in the estimate of critical parameters, we introduce a new estimate $\hat{\alpha}_c^{\overline{y_{\alpha, N}}}$ obtained by removing the datapoint $y_{\alpha, N}$ from the dataset (whose size we denote by $D$). Denoting the number of datapoints by $D$, the variance in the estimate is taken to be
\begin{equation}
       \text{Var}(\alpha_c) = \frac{D-1}{D}\sum_{y_{\alpha, N}} \left(\overline{\alpha_c}-\hat{\alpha}_c^{\overline{y_{\alpha, N}}}\right)^2
\qquad {\text{where} } \quad 
    \overline{\alpha_c} = \frac{1}{D}\sum_{y_{\alpha, N}} \hat{\alpha}_c^{\overline{y_{\alpha, N}}},
\end{equation}

The error in $\nu$ is obtained similarly.

\section{Bootstrap Estimate of Error}
\label{app:boostrapping-estimate}
The bootstrap resampling technique is commonly used to estimate errorbars and confidence intervals when straightforward error propagation is difficult. We use this procedure to estimate the errorbars for experimental data. Given an sample of size $N$, we generate 1000 new samples of size 20. This is done by uniformly picking elements from the original sample with replacement. We take the standard deviation of the means of the new samples to be the boostrap error. 

\section{Analytics on R\'enyi-approximation to Conditional Entropy}
\label{app:renyi-analogue-analytics}
In this section, we use Schur-Weyl duality to analytically approximate the R\'enyi analogue of the conditional entropy, $S_{\ell|s}^{(2)} = S_{\ell,s}^{(2)}-S_{s}^{(2)}$, with $S_{X}^{(2)} = -\log \sum_{x \in X} p_x^2$ being the R\'enyi entropy. As a reminder, $\ell$ is the distribution of measurements of the $K$ logical qubits, and $s$ is the distribution of measurements of the $N-K$ syndrome qubits, both in the computational basis. We would like to compute the \Renyi, analogue of the conditional entropy, averaged over Clifford encoders.  

The circuit-averaged measure we are interested in is therefore,
\begin{equation}
   \mathbb{E}_C S_{\ell|s}^{(2)} = \mathbb{E}_C S_{\ell, s}^{(2)} -   \mathbb{E}_C S_s^{(2)} =- (\mathbb{E}_C \log M_N -   \mathbb{E}_C \log M_{N-K}) \qquad {\text{with   }} M_k = \sum_{x \in \{0,1\}^{k}} p_x^2
\end{equation}
where we have introduced the notation $M_k$ to denote the collision probability over the distribution of measurements of $k$ qubits. 

Crucially, if the circuit-to-circuit variability of the collision probability is negligible, we can take the expectation over Clifford circuits inside the logarithm. We observe that such a typicality assumption is indeed valid for sufficiently large system, but breaks down for small $N$ and large error-rate $\alpha \to \pi/2$. Here, we proceed with the typicality assumption to get
\begin{equation}
     \mathbb{E}_C S_{\ell|s}^{(2)} \approx - (\log \mathbb{E}_C  M_N -   \log  \mathbb{E}_C M_{N-K})
     \label{eq:circuit-averaged-renyi}
\end{equation}

Consider the circuit-averaged collision probability in our model model where we begin with qubits in a $\ket{0}^{\otimes N}$ state, followed by a Clifford unitary $C$, followed by a noise operation $N(\alpha) = \prod_{i=1}^N \exp(i\sigma_z \alpha/2)$, followed by $C^\dagger$. We finally measure all $N-K$ qubits at the end. distribution induced by measurements of $N-K$ qubits out of a $N$ qubit system $\ket{\psi}$:
\begin{align*}
    \mathbb{E}_C M_{N-K} &=  \sum_{x \in \{0,1\}^{N-K}} p_x^2 \\
    &= \sum_{x \in \{0,1\}^{N-K}}  \tr\left(I_{2^K} \otimes \ket{x}\bra{x} \ket{\psi}\bra{\psi} \right)^2
    \\
    &=\sum_{x \in \{0,1\}^{N-K}} \mathbb{E}_C \tr\left((I_{2^K} \otimes \ket{x}\bra{x}) C^\dagger N(\alpha) C \ket{0}\bra{0} C^\dagger N(\alpha)^\dagger C \right)^2  \\
    &= \sum_{x \in \{0,1\}^{N-K}} \mathbb{E}_C \tr\left((I_{2^K} \otimes \ket{x}\bra{x}) \otimes  (I_{2^K} \otimes \ket{x}\bra{x})) \left(C^\dagger\right)^{\otimes 2} \right. \\ & \quad \quad\quad  \left. N(\alpha)^{\otimes 2} C^{\otimes 2} \ket{0,0}\bra{0,0} \left(C^\dagger\right)^{\otimes 2} \left(N(\alpha)^\dagger\right)^{\otimes 2} C^{\otimes 2} \right), 
\end{align*}
where we define $I_{2^K}$ to be the identity operator on $K$ logical qubits.  Substituting the basis-decomposed representation of $I_{2^K} = \sum_{i \in \{0, 1\}^K} \ket{i}\bra{i}$ above, we get
{\small \begin{align*}
    & \mathbb{E}_C M_{N-K} \\ 
    &= \sum_{\substack{x\in \{0,1\}^{N-K} \\ i,j \in \{0,1\}^{K} }}  \mathbb{E}_C \tr\left((\ket{i, x}\bra{i, x}) \otimes  (\ket{j, x}\bra{j, x})) \right. \\ & \qquad  \left. \hspace{80pt}\left(C^\dagger\right)^{\otimes 2} N(\alpha)^{\otimes 2} C^{\otimes 2} \ket{0^N,0^N}\bra{0^N,0^N} \left(C^\dagger\right)^{\otimes 2} \left(N(\alpha)^\dagger\right)^{\otimes 2} C^{\otimes 2} \right)  \\
    &= \sum_{\substack{x\in \{0,1\}^{N-K} \\ i,j \in \{0,1\}^{K} }}
    \mathbb{E}_C \left[\bra{(i, x), (j,x)} \left(C^\dagger\right)^{\otimes 2} \mathcal{N}(\alpha)^{\otimes 2} C^{\otimes 2} \ket{0^n,0^n}\right. 
 \\& \hspace{100pt}\left. \bra{0^N,0^N} \left(C^\dagger\right)^{\otimes 2} \left(\mathcal{N}(\alpha)^\dagger\right)^{\otimes 2} C^{\otimes 2} \ket{(i, x), (j,x)}\right] \\
    &= \sum_{\substack{x\in \{0,1\}^{N-K} \\ i,j \in \{0,1\}^{K} }} \mathbb{E}_C \braket{(i, x), (j,x), 0^n, 0^n |  \left(C^\dagger\right)^{\otimes 4} \left( \mathcal{N}(\alpha)^{\otimes 2}\otimes (\mathcal{N}(\alpha)^\dagger)^{\otimes 2}\right) C^{\otimes 4}|0^n, 0^n, (i, x), (j,x)}\\
    &= \sum_{\substack{x\in \{0,1\}^{N-K} \\ i,j \in \{0,1\}^{K} }} \braket{(i, x), (j,x), 0^N, 0^N |  \mathbb{E}_C \left[ \left(C^\dagger\right)^{\otimes 4} \left( \mathcal{N}(\alpha)^{\otimes 2}\otimes (\mathcal{N}(\alpha)^\dagger)^{\otimes 2}\right) C^{\otimes 4}\right]|0^N, 0^N, (i, x), (j,x)} 
    \label{eq:fourth-moment-C}
\end{align*}
}
\noindent In the second line, we have used the relation $\tr(\ket{x}\bra{x} A \ket{0}\bra{0} B) = \braket{x|A|0}\braket{0|B|x}$. In the third line, we combine the two inner products  $ \braket{x|A|0}\braket{0|B|x} = \braket{x,0|(A \otimes B)|0,x}$. In the fourth line, we have moved the expectation inside the inner-product. 

The Schur-Weyl duality \cite{gross2021schur} gives a decomposition of a Clifford-averaged operator as a linear sum of representation of a semigroup $\Sigma_4$ which consists of 30 elements.\begin{equation}
    \mathbb{E}_C \left[ \left(C^\dagger\right)^{\otimes 4} \left( N(\alpha)^{\otimes 2}\otimes (N(\alpha)^\dagger)^{\otimes 2}\right) C^{\otimes 4}\right] = \sum_{T \in \Sigma_{4}}a_T R(T)
\end{equation}
 $R(T)$ is the representation for $T \in \Sigma_4$ operator which acts on four copies of the $N$-qubit state, and $a(T)$ is the corresponding weight. Knowing all the representations $R(T)$, it is possible to calculate the coefficients $a_T$ for each error-rate $\alpha$ \cite{gross2021schur}. One way of doing so is by solving the set of $30$ simultaneous equations: $\forall S \in \Sigma, \tr{[R(S)O]} = \sum_{T}a_T \tr{[R(S)R(T)]}$, where $\Sigma$ are the generators of the semigroup corresponding to the Clifford group, and $O$ is the operator we would like to decompose (in our case, $O = \left( N(\alpha)^{\otimes 2}\otimes (N(\alpha)^\dagger)^{\otimes 2}\right)$). Substituting the decomposition into our expression for $\mathbb{E}_C M$, we get,
\begin{equation}
    \mathbb{E}_{C}M_{N-K}   = \sum_{\substack{x\in \{0,1\}^{N-K} \\ {i,j \in \{0,1\}^{K}}}} \sum_{T \in \Sigma_4} a_T \braket{(i, x), (j,x), 0^N, 0^N |  R(T)|0^N, 0^N, (i, x), (j,x)} 
\end{equation}
Note that $R(T)$ is a qubit-wise representation, that is $R(T) = r(T)^n$ for some $r(T)$ acting on four copies of a single-qubit state.  We can then re-write the expression above, distributing the representation to the "syndrome" register and the "logical qubit" register. 
\begin{equation}
    \mathbb{E}_C M_{N-K}   = \sum_{\substack{x\in \{0,1\}^{N-K} \\ {i,j \in \{0,1\}^{K}}}}  \sum_{T \in \Sigma_4} a_T \braket{i,j,0^K,0^K|r(T)^{K}|0^K,0^K,i,j}\braket{x,x,0^N,0^N| r(T)^{n-k}|0^N, 0^N, x,x} 
\end{equation}
We would like to evaluate this expression. First, consider the case where $x=0$. The second inner-product resolves to $1$ for all $T$ when $x=0$. The first inner product also resolves to $1$ for all $T$ if $i=j=0$. If $i=0$ but $j \neq 0$, a certain subset of $\Sigma_4$ resolve to one (the rest evaluate to zero). Let's call this set $S_{0,x}$). Similarly, if $j=0$ and $i\neq 0$, denote the subset that resolves to $1$ by $S_{x,0}$). Similarly, let the set $S_{x,x}$ denote elements that evaluate to $1$ whenever $i=j\neq 0$. Finally, let $S_{x,y}$ be the set of elements that resolves $\braket{i,j,0,0|r(T)^K|0,0,i,j}$ to one whenenver $i\neq 0, j\neq 0, i\neq j$. The total contribution of the $x=0$ term is given by
\begin{align}
\begin{split}
  & {\sum_{T \in \Sigma_4} a_T}_{i=0,j=0} + \underbrace{\sum_{i \in \{0, 1\}^{K}} \sum_{T \in S_{x,0}}a_T}_{i\neq 0, j=0} + \underbrace{\sum_{\substack{j\in \{0, 1\}^{K} \\ j \neq 0}} \sum_{T \in S_{0,x}}a_T}_{i=0,j \neq 0} + \underbrace{\sum_{\substack{j\in \{0, 1\}^{K} \\ j \neq 0}} \sum_{T \in S_{x,x}}a_T}_{i=j\neq 0} + \sum_{\substack{i,j\in \{0, 1\}^{K}\\ i\neq j \\ i\neq 0, j\neq 0}} \sum_{T \in S_{x,x}}a_T \\\ 
  &= \sum_{T \in \Sigma_4} a_T + (2^K-1) \sum_{T \in S_{x,0}}a_T+ (2^K-1) \sum_{T \in S_{0,K}}a_T \\
  &\quad+ (2^K-1) \sum_{T \in S_{x,x}}a_T + (2^K-1)(2^K-2) \sum_{T \in S_{x,y}} a_T
  \label{eq:First-half}
  \end{split}
\end{align}
Second, whenever $x \neq 0$, the second inner product resolves to $1$ only for the set $S_{x,x}$, otherwise it evaluates to 0. The condition that $T \in S_{x,x}$ also necessitates that $i=j$ in the first inner product. The set $\{x \neq 0\}$ has size $2^{N-K}-1$. Therefore, the total contribution from the $x \neq 0$ terms is simply
\begin{equation}
    (2^{N-K}-1) \left( 2^K \sum_{T \in S_{x,x}}a_T  + 2^K (2^K-1) \sum_{T \in S_{x,y}}a_T\right) 
    \label{eq:second-half}
\end{equation}
Adding \eqref{eq:First-half} and \eqref{eq:second-half}, we get the expected $M_K$. 

\medskip 

Finally, we approximate the the circuit-averaged Renyi-analogue of the conditional entropy using \eqref{eq:circuit-averaged-renyi}.

\section{$r \to 1$ Limit}
{ In the main text we focus on numerics for $r \leq 1/2$. But the $r \to 1$ limit is interesting in that that it is analytically solvable, because it has no syndrome measurements. 

To proceed, we note that since, magic is invariant under Clifford operations, the magic corresponding to a circuit $C$ is given by,
\begin{equation}
  \mathcal{M}_C \equiv \mathcal{M} [C \mathcal{N}(\alpha)^\dagger C^\dagger \rho_0 C \mathcal{N}^{\dagger}(\alpha) C^\dagger] = \mathcal{M} [\mathcal{N}^\dagger(\alpha) C^\dagger \rho_0 C \mathcal{N}(\alpha)]
\end{equation}
The Clifford-averaged magic is therefore
\begin{align}
    \langle  \mathcal{M}_C \rangle = \mathbb{E}_C \left[ -\log \frac{1}{2^N}\sum_{P \in \mathcal{P}_N} \text{Tr}( \mathcal{N}^\dagger(\alpha) C^\dagger \rho_0 C \mathcal{N}(\alpha) P)^4 \right] \approx -\log \frac{1}{2^N}\sum_{P \in \mathcal{P}} 
 \mathbb{E}_C \left[ \text{Tr}(\mathcal{N}^\dagger(\alpha) C^\dagger \rho_0 C \mathcal{N}(\alpha) P)^4  \right],
\end{align}
where we have used a typicality argument to move the expectation inside the logarithm. Now we can calculate the expectation of the fourth moments of Pauli operators
\begin{align}
    \mathbb{E}_C \left[ \text{Tr}\left( \mathcal{N}^\dagger(\alpha)C^\dagger \rho_0 C N(\alpha) P\right)^4  \right]  &=   \mathbb{E}_C \left[ \text{Tr}(N^\dagger(\alpha) C^\dagger \rho_0 C N(\alpha) P)^4  \right] \\
    &= \mathbb{E}_C \left[ \text{Tr}\left(\mathcal{N}^\dagger(\alpha)^{\otimes 4} \left( C^\dagger\right)^{\otimes 4} \rho_0^{\otimes 4} C^{\otimes 4} \mathcal{N}(\alpha)^{\otimes 4} P^{\otimes 4}\right) \right] \\
    &= \text{Tr}\left(\mathcal{N}^\dagger(\alpha)^{\otimes 4} \mathbb{E}_C \left[ \left( C^\dagger\right)^{\otimes 4} \rho_0^{\otimes 4} C^{\otimes 4}\right) \mathcal{N}(\alpha)^{\otimes 4} P^{\otimes 4} \right)
\end{align}
Averaging over Clifford unitaries, using techniques from \cite{gross2021schur}, gives us:
\begin{equation}
   \mathbb{E}_C \left[ \left( C^\dagger\right)^{\otimes 4} \rho_0^{\otimes 4} C^{\otimes 4}\right) =  \frac{1}{Z} \sum_{T \in \Sigma_{4}} R(T) =  \frac{1}{Z} \sum_{T \in \Sigma_{4}} r(T)^{\otimes N},
\end{equation}
where $Z$ is the normalizing constant $Z_N = 2^N \left(2^N+1\right) \left(2^N+2\right) \left(2^N+4\right)$ and $R(T)$ is the representation for $T \in \Sigma_4$ operator which acts on four copies of the $N$-qubit state
Using it together with the sum over all Pauli operators over $n$ qubits
\begin{align}
 \sum_{P \in \mathcal{P}_N} \mathbb{E}_C \left[ \text{Tr}\left( \mathcal{N}^\dagger(\alpha)C^\dagger \rho_0 C \mathcal{N}(\alpha) P\right)^4  \right] &=  \sum_{P \in \mathcal{P}_N} \mathbb{E}_C \left[ \text{Tr}\left( \mathcal{N}^\dagger(\alpha)C^\dagger \rho_0 C \mathcal{N}(\alpha) P\right)^4  \right] \\
 &= \frac{1}{Z_n} \sum_{T \in \Sigma_{4}} \sum_{P \in \mathcal{P}_n} \text{Tr}\left(\mathcal{N}^\dagger(\alpha)^{\otimes 4}r(T)^{\otimes N} \mathcal{N}(\alpha)^{\otimes 4} P^{\otimes 4} \right)
\end{align}
Noting that $\sum_{P \in \mathcal{P}_N} P^{\otimes 4} = \left(\sum_{P \in P_1} P^{\otimes 4} \right)^{\otimes N}$, we can rewrite the above as
\begin{align}
    \frac{1}{Z_N} \sum_{T \in \Sigma_{4}}  \text{Tr}\left(\mathcal{N}^\dagger(\alpha)^{\otimes 4}r(T)^{\otimes N} \mathcal{N}(\alpha)^{\otimes 4} P^{\otimes 4} \right) &= \frac{1}{Z_N} \sum_{T \in \Sigma_{4}}  \text{Tr}\left(\text{Rz}^\dagger(\alpha)^{\otimes 4n}r(T)^{\otimes n} \text{Rz}(\alpha)^{\otimes 4N}  \left(\sum_{P \in P_1} P^{\otimes 4} \right)^{\otimes N} \right) \\
    &=  \frac{1}{Z_N} \sum_{T \in \Sigma_{4}}  \text{Tr}\left(\text{Rz}^\dagger(\alpha)^{\otimes 4}r(T) \text{Rz}(\alpha)^{\otimes 4}  \left(\sum_{P \in P_1} P^{\otimes 4} \right) \right)^{ N},
\end{align}
Taking the logarithm of above gives the Clifford-averaged magic
\begin{equation}
  \langle  \mathcal{M}\rangle  = N + \log(Z_N) - \log  \left(\sum_{T \in \Sigma_{4}}  \text{Tr}\left(\text{Rz}^\dagger(\alpha)^{\otimes 4}r(T) \text{Rz}(\alpha)^{\otimes 4}  \left(\sum_{P \in P_1} P^{\otimes 4} \right) \right)^{ N}\right)
\end{equation}
In the limit $N \to \infty$, the magic density remains a constant, implying a volume-law scaling, regardless of the size:
\begin{equation}
    \lim_{N\to \infty} \frac{1}{N}\langle  \mathcal{M}\rangle = -\log\left[1 - \frac{\sin^2(2 \alpha )}{4}\right].
\end{equation}
This analytic limit, together with numerically calculated SSRE for tractable sizes, is presented in Fig.~\ref{fig:large-r}.
\begin{figure}[!ht]
    \centering
    \includegraphics[width=0.5\textwidth]{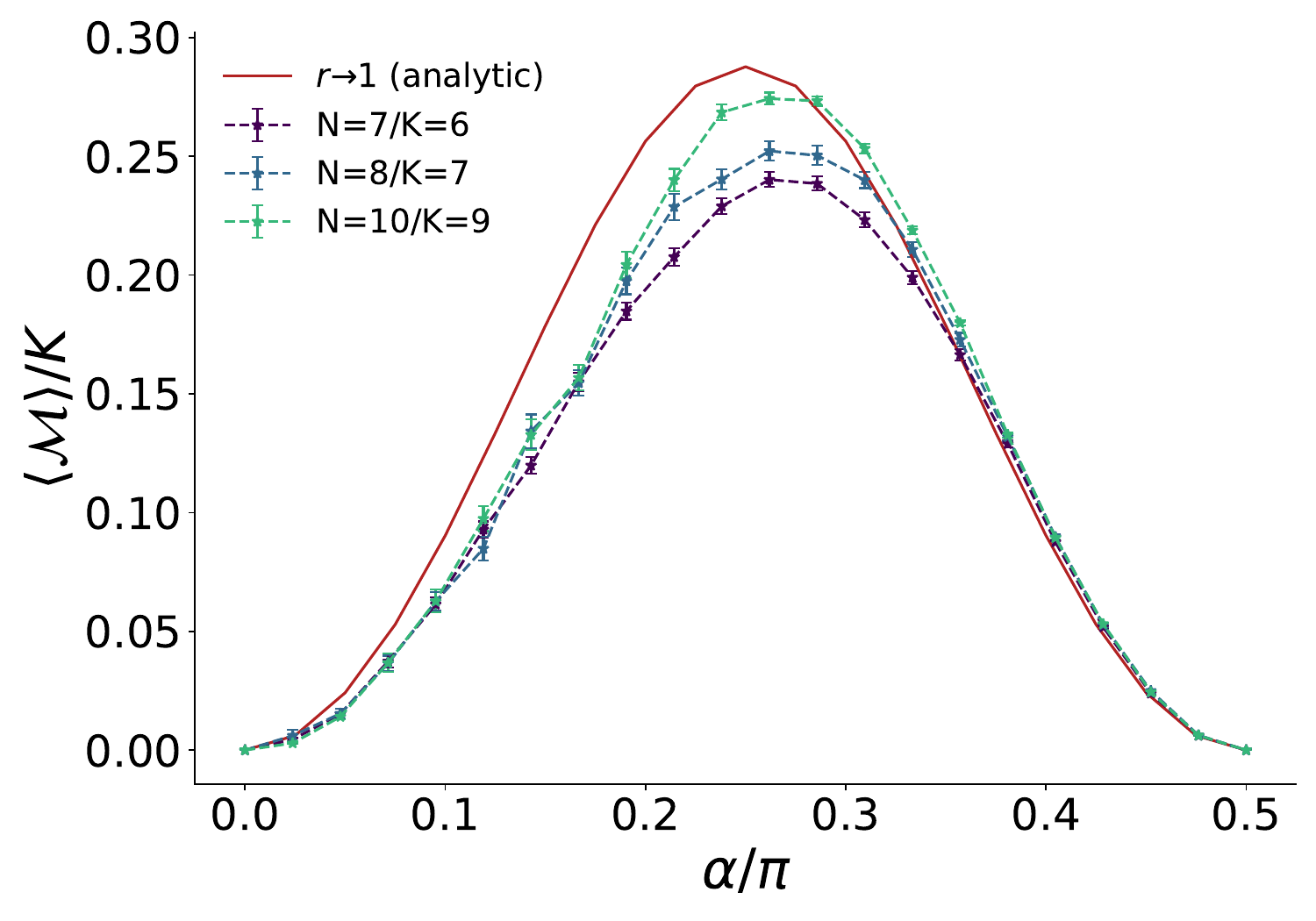}
    \caption{Numerically evaluated SSRE for $r = K/N$, close to 1 (dashed lines) and the analytical $r \to 1$ limit (solid red). The behavior of SSRE is consistent with what we observe with lower rates.}
    \label{fig:large-r}
\end{figure}

}

\section{Noisy Simulation}
\begin{figure}[!ht]
\centering\includegraphics[width=0.5\textwidth]{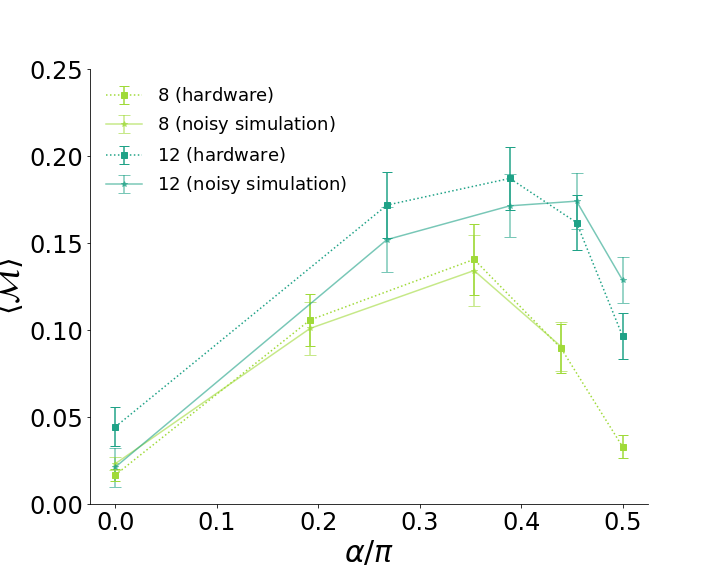}
    \caption{Comparison of hardware data with noisy simulation for vanishing-rate magic experiments with $L=8, 12$. The noise model applies overrotation and depolarization to $XX$ gates.}
    \label{fig:noisy_simulation}
\end{figure}

For noisy simulation, we consider a noise model where every two-qubit XX gate is followed by a noise model that consists of XX overrortation followed by a depolarizing channel.
\begin{equation}
    \exp\left(i\frac{\pi}{2}\text{XX}\right) \to \exp\left(i\frac{\pi}{2}\text{XX}(1 + \epsilon)\right) \to (1-p)\exp\left(i\frac{\pi}{2}\text{XX}(1 + \epsilon)\right) + \frac{p}{4}\mathbb{1}.
\end{equation}
The over-rotation noise, $\epsilon$, is sampled from a Gaussian distribution $\mathcal{N}(0, 0.06)$ independently for each gate and each shot, and the depolarizing noise rate $p = 0.016$. In Fig.~\ref{fig:noisy_simulation}, we compare noisy simulations with experiments for vanishing rate magic experiments with $L=8$ and $12$.

\section{Data Without Post-Processing}
\begin{figure}
    \centering
    \includegraphics[width=\textwidth]{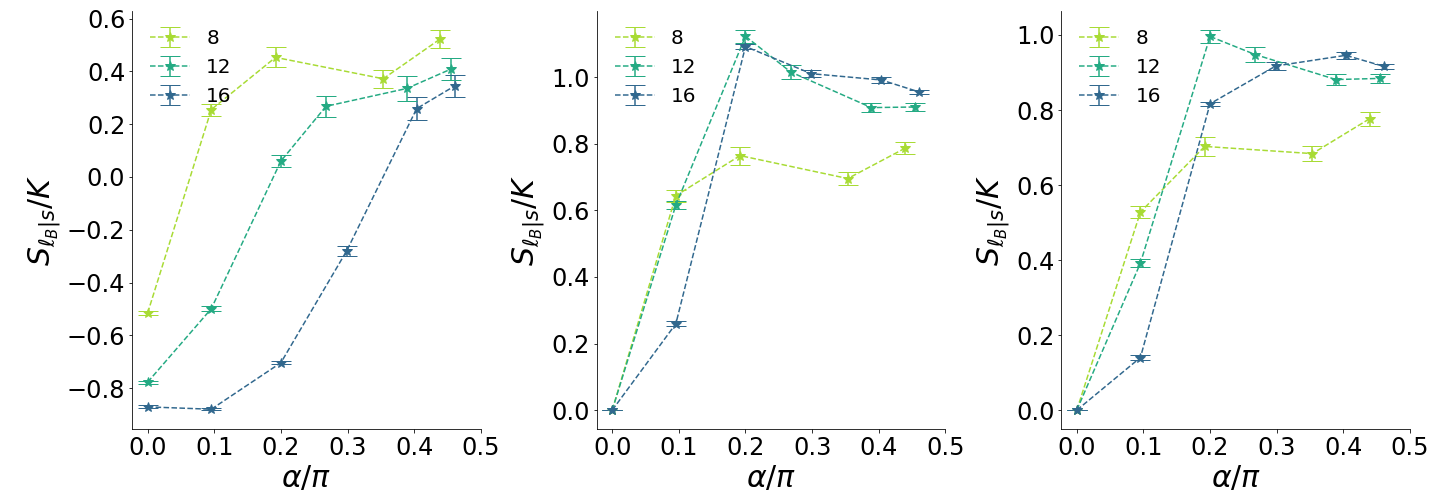}
    \caption{Conditional Shannon entropy with varying levels of error mitigation. Left: no error mitigation  Center: The experimental data is projected onto the ideal subspace, and we compute $\sum_{x \in \rm Ideal} p(x)\log p(x)$ where $p(x)$ is the experimentally measured probability. Right) The experimental data is projected onto the ideal subspace, and we compute $\sum_{x \in \rm Ideal} p(x)\log \tilde{p}(x)$ where $p(x)$ and $\tilde{p}(x)$ are the experimentally measured and theoretically calculated probabilities respectively. }
    \label{fig:unmitigated_shannon}
\end{figure}

\begin{figure}
    \centering
    \includegraphics[width=\textwidth]{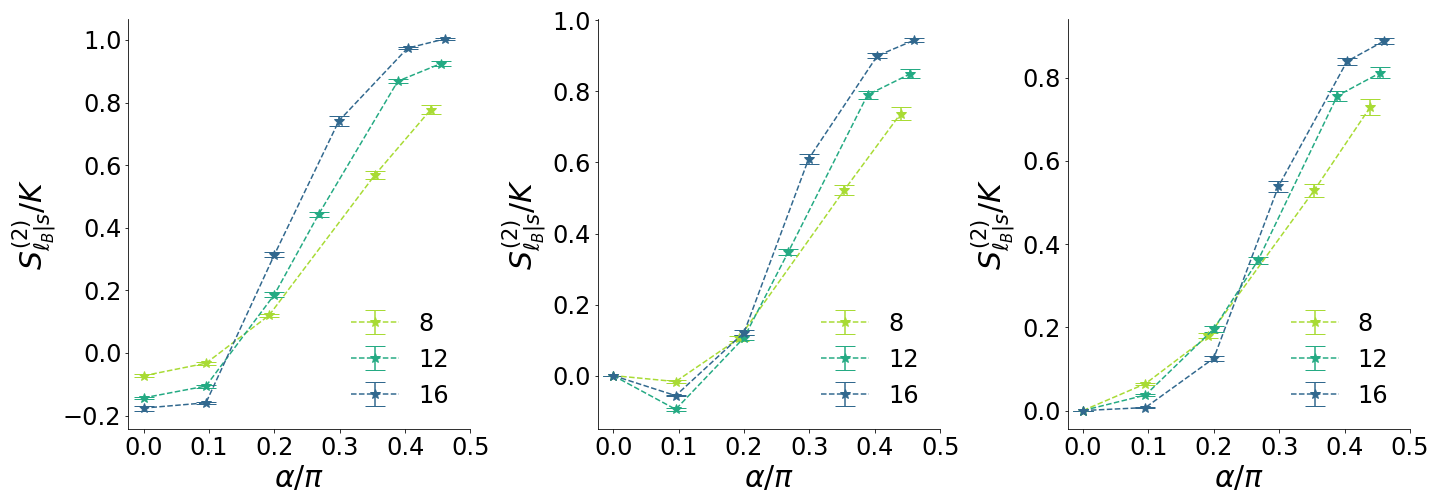}
    \caption{Renyi entropy with varying levels of error mitigation. Left: no error mitigation  Center: The experimental data is projected onto the ideal subspace, and we compute $\sum_{x \in \rm Ideal} p(x)^2$ where $p(x)$ is the experimentally measured probability. Right) The experimental data is projected onto the ideal subspace, and we compute $\sum_{x \in \rm Ideal} p(x) \tilde{p}(x)$ where $p(x)$ and $\tilde{p}(x)$ are the experimentally measured and theoretically calculated probabilities respectively. }
    \label{fig:unmitigated_renyi}
\end{figure}

The constant-rate experimental data presented  has undergone post-processing, where we apply error-mitigation. In particular, we project the experimentally obtained probability distribution to the probability distribution of the ideal (noiseless) wavefunction. In Fig.~\ref{fig:unmitigated_shannon} (Fig.~\ref{fig:unmitigated_renyi}), we present the conditional Shannon (Renyi) entropy with varying degrees of post-processing.  

\section{Constrained Scaling Collapse of Constant-Rate Magic Phase Transition}
\label{sec:constrained-gamma}
{
We note in the main text that, in the scaling collapse for constant rate magic in Fig.~\ref{fig:constant-rate-magic-full-panel}A, the $\gamma\approx 1.2$ exponent we obtain via free scaling collapse in unphysical. We believe this to be the artifact of the small system sizes we have data for. Between area law and volume law, we expect the scaling dimension of magic ought to be bounded between $[0, 1]$. Here, we present the scaling collapse where we constrain $\gamma$ to this range of physical value.
\begin{figure}
    \centering
    \includegraphics[width=0.5\textwidth]{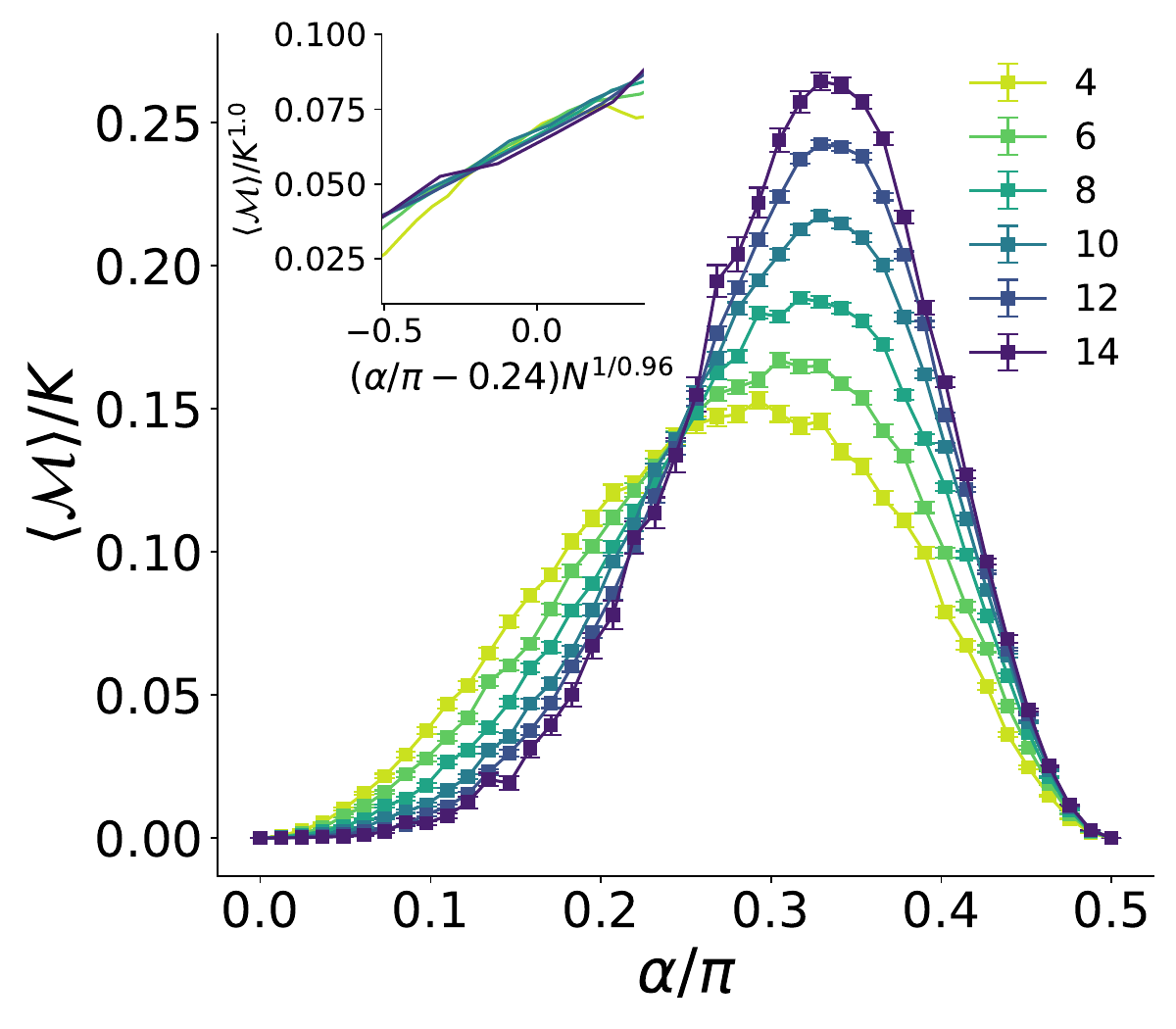}
    \caption{Density of magic (SSRE)  of the logical space and its scaling collapse (inset) plotted against the error rate $\alpha$, for code rate $r = K/N = 1/2$. The error bars are derived using standard error and are omitted in the scaling collapse (inset), where the x-axis is scaled as $(\alpha/\pi-\alpha_c)N^{1/\nu}$ with critical parameters $\alpha_c = 0.240(2)$ and $\nu = 0.96(5)$, and the y-axis is scaled as $\langle \mathcal{M}\rangle/K^{\gamma}$ with $\gamma = 1.000$.}
    \label{fig:constant-rate-constrained}
\end{figure}

}
\end{document}